\documentclass[epjST]{svjour}
\usepackage{graphics}
\usepackage{mwe}
\usepackage{color}
\usepackage{xspace}
\usepackage{amsmath}
\usepackage[colorlinks=true]{hyperref}
\begin{document}
\title{The \neut Neutrino Interaction Simulation Program Library}

\author{Yoshinari Hayato\inst{1}\fnmsep\thanks{\email{hayato@suketto.icrr.u-tokyo.ac.jp}} \and Luke Pickering\inst{2}}
\institute{The University of Tokyo, ICRR, Kamioka Observatory and Kavli IPMU, Kamioka, Gifu, Japan \and Michigan State University, Department of Physics and Astronomy,  East Lansing, Michigan, U.S.A.}
\newcommand{\neut}{\texttt{NEUT}\xspace}
\newcommand{\cernlib}{\texttt{CERNLIB}\xspace}
\newcommand{\sk}{SK\xspace}
\newcommand{\ttwok}{T2K\xspace}
\newcommand{\erecqe}[0]{\ensuremath{E^{\textrm{Rec}}_{\textsc{qe}}}\xspace}

\abstract{
\neut is a neutrino-nucleus interaction simulation program library. It can be used to simulate interactions for neutrinos with between 100~MeV and a few TeV of energy. \neut is also capable of simulating hadron interactions within a nucleus and is used to model nucleon decay and hadron--nucleus
interactions for particle propagation in detector simulations. This article describes the range of interactions modelled and how each is implemented.
} 
\maketitle

\section{Introduction}\label{sec:Introduction}
\neut is primarily a neutrino--nucleus scattering simulation program library and provides a complete model capable of predicting the observations for a wide range of neutrino scattering experiments. \neut is capable of simulating neutrino--nucleon and coherent neutrino--nucleus interactions in a number of reaction channels over a neutrino energy range from 100 MeV to a few TeV. Additionally, \neut incorporates initial- and final state nuclear effects for interactions with nuclei from boron to lead. One of the most important nuclear effects is the re-scattering of hadrons, which are produced in the primary neutrino--nucleon interaction, as they propagate out of the nuclear medium. This re-scattering can result in hadron absorption, extra hadron production or knock-out, or distortion of the nuclear-leaving particle kinematic spectra. The \neut hadron re-scattering model has also been used to simulate low-energy pion--nucleus scattering both to tune the model to extant data~\cite{Elder:2019} and to simulate pion propagation in neutrino-scattering experimental simulations. Finally, \neut can also simulate various nucleon decay channels to support experimental searches for the process.

\neut has a long rich history, originally developed in the 1980s as a tool to study atmospheric neutrinos and nucleon decay in the Kamiokande experiment~\cite{Kamiokande:1986iox}, and some of the original \texttt{FORTRAN}77 code is still in use. \neut continues to be predominantly developed and maintained by members of the Kamiokande series of experiments (Super-Kamiokande, T2K, Hyper-Kamiokande) and many source files contain comments messages from the numerous physicists who have contributed to the simulation over the past 35 years---including those working on the Nobel prize-winning Super-Kamiokande (SK) analysis~\cite{Super-Kamiokande:1998kpq}. Recent development has targeted the improvements most-needed for precise and robust analyses of \sk and \ttwok neutrino oscillation data and neutrino cross-section measurements. At \ttwok energies, Charged-current quasi-elastic interactions dominate. It has become clear over the last decade that such interactions can only be precisely predicted by incorporating detailed models of relevant nuclear dynamics. For the multi-GeV samples used in \sk analyses, shallow and deep inelastic scattering channels are critical for modelling the expected rate of multi-ring events seen in the detector. The transition region between resonance excitation and deep inelastic scattering has proven particularly difficult to model well. 
Because of the in-house nature of \neut development and analysis usage, it is not yet open source. We do not yet have the resources to migrate to an open source model, but, access to the code and usage instructions are available upon request.

\section{Motivating the use of an Interaction Simulation}\label{sec:Structure}

\begin{figure}
  \centering
  \includegraphics[width=0.6\textwidth]{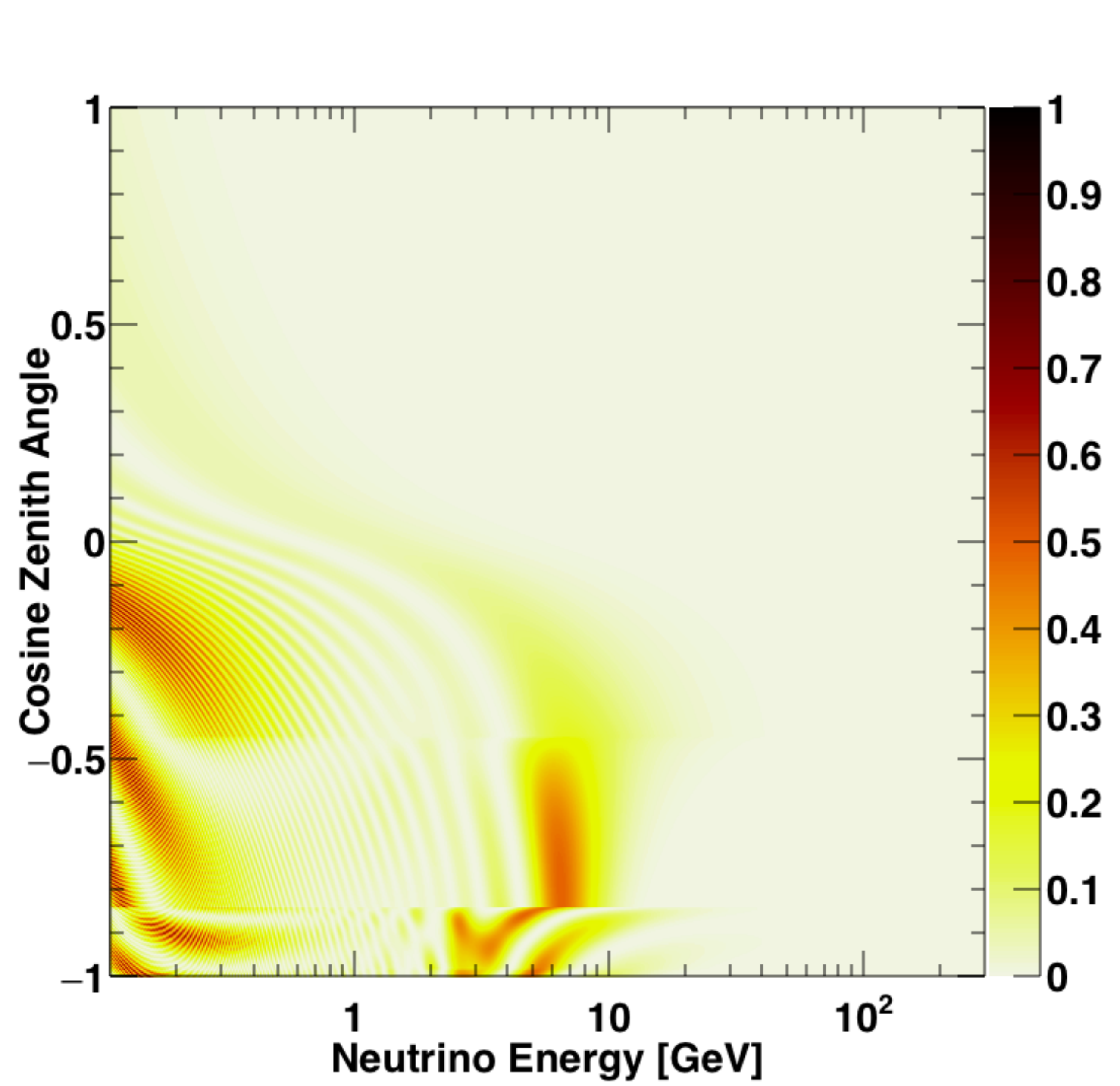}
  \includegraphics[width=0.6\textwidth]{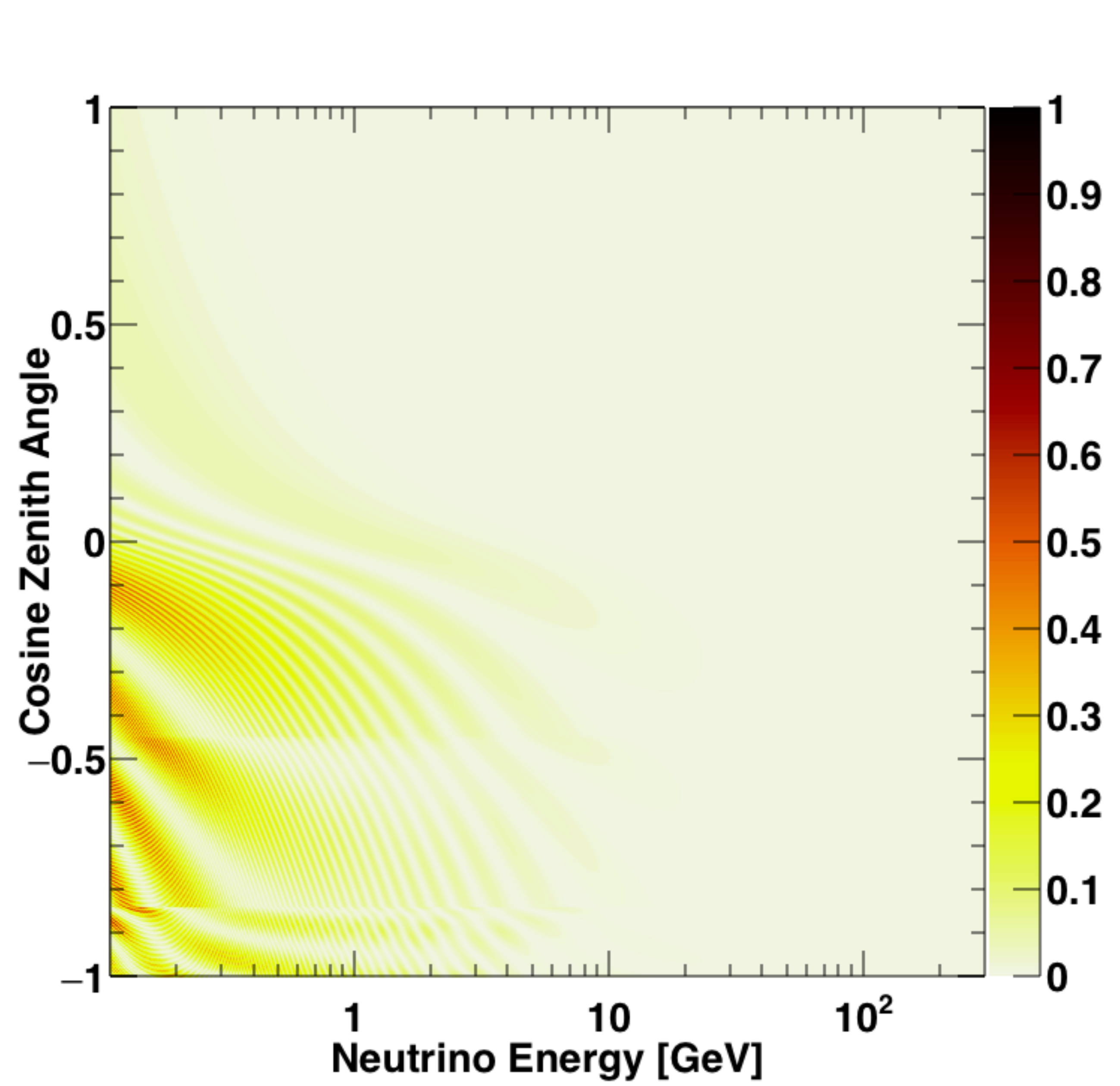}
  \caption{Oscillation probabilities for atmospheric $\nu_\mu$ to $\nu_e$ (top)
  and $\bar{\nu}_\mu$ to $\bar{\nu}_e$ (bottom) as a function of neutrino energy and the cosine of the
  zenith angle (acting as a proxy for propagation distance through the Earth) assuming a normal neutrino mass hierarchy. Matter effects in the Earth 
  produce distortions in oscillations of $\nu_\mu$ between two and ten~GeV, 
  which are not seen for $\bar{\nu}_\mu$ oscillations. Figures reproduced from Ref.~~\cite{Super-Kamiokande:2017yvm}
  The oscillation parameters used are: 
            $\Delta m_{32}^{2} = 2.5 \times 10^{-3} \mbox{eV}^{2}$, 
            $\mbox{sin}^{2} \theta_{23} = 0.5$, 
            $\mbox{sin}^{2} \theta_{13} = 0.0219$, and
            $\delta_{CP} = 0$.}
  \label{fig:Introduction:ATMNU_Oscillogram}
\end{figure}

\begin{figure}
  \centering
  \includegraphics[width=0.7\textwidth]{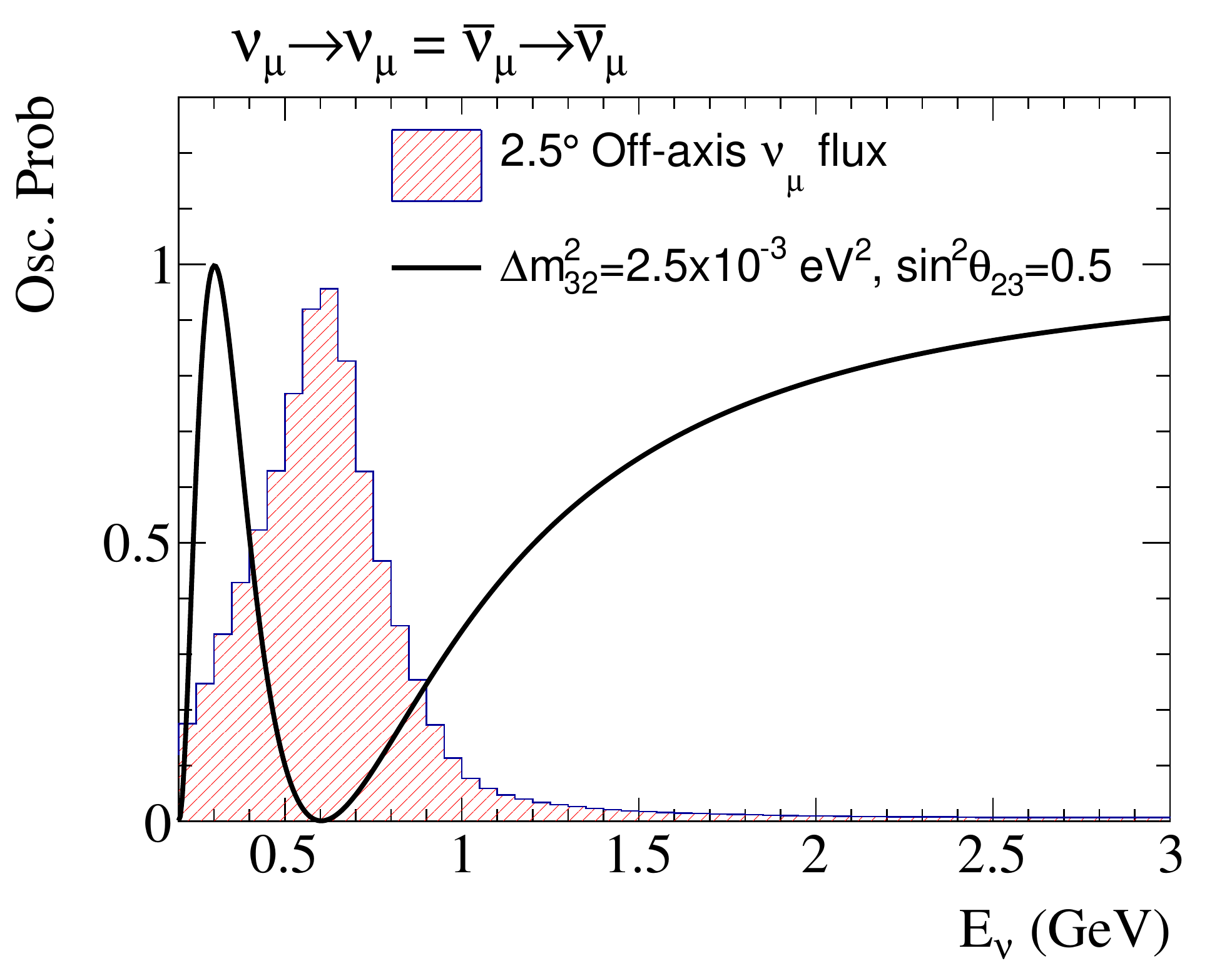}
  \includegraphics[width=0.7\textwidth]{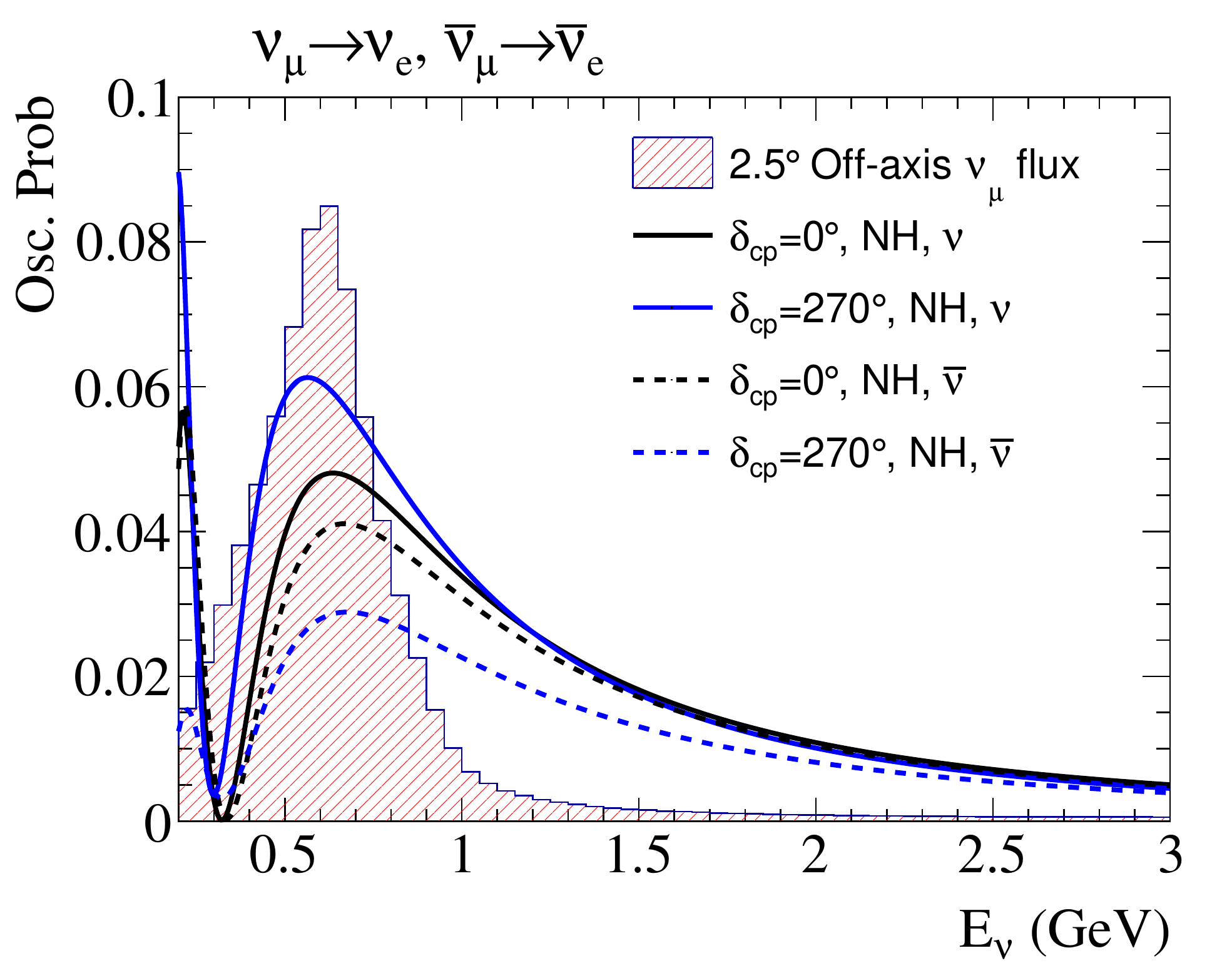}
  \caption{The T2K $\nu_\mu$ flux~\cite{T2K:2015sqm,T2K:2012nrr} (energy spectral shape) 
  and overlaid oscillation
  probabilities for the $\nu_\mu$ to $\nu_\mu$ (top)
  and $\nu_\mu$ to $\nu_e$ and $\nu_\mu$ to $\nu_e$ (bottom) as a function of neutrino energy.
    The black (blue) curves correspond to minimal (maximal) values of the CP-violating phase $\delta_\textsc{cp}$.
  The oscillation parameters used are:
  $\Delta m^2_{32}=2.5\times10^{-3}\rm{eV^2}$,
  $\sin^2(\theta_{23})=0.5$,
  $\sin^2(2\theta_{13})=0.1$.
}
  \label{fig:Introduction:T2K_OscProb}
\end{figure}

The primary goal of atmospheric and long-baseline neutrino-scattering experiments is to study neutrino oscillation, which occurs as a function of neutrino energy and flavor. As neutrinos are neutral particles, their properties (including their energy) can only be inferred from observable secondary particles produced when they interact with matter in our particle detectors.
Interaction simulations are used to predict the probability of neutrinos of a given flavor and energy to interact with a given target, the observable secondary particle spectra produced in the neutrino--nucleus interaction, and event selection efficiencies and purities. 
The common use of nuclear targets in neutrino scattering experiments further complicates the problem as secondary hadrons produced in the \emph{primary} neutrino interaction can be absorbed or lose energy before leaving the target nucleus, obfuscating the details of the primary interaction. As a result, interaction simulations are a critical tool in the analysis of neutrino oscillation data. The neutrino energy range modeled most carefully by \neut is between 0.1 and 10~GeV. This range is motivated by the typical energies for oscillation features expected in \sk and \ttwok data, as seen in Fig.~\ref{fig:Introduction:ATMNU_Oscillogram} and Fig.~\ref{fig:Introduction:T2K_OscProb}.

\begin{figure}
  \centering
  \includegraphics[width=0.8\textwidth]{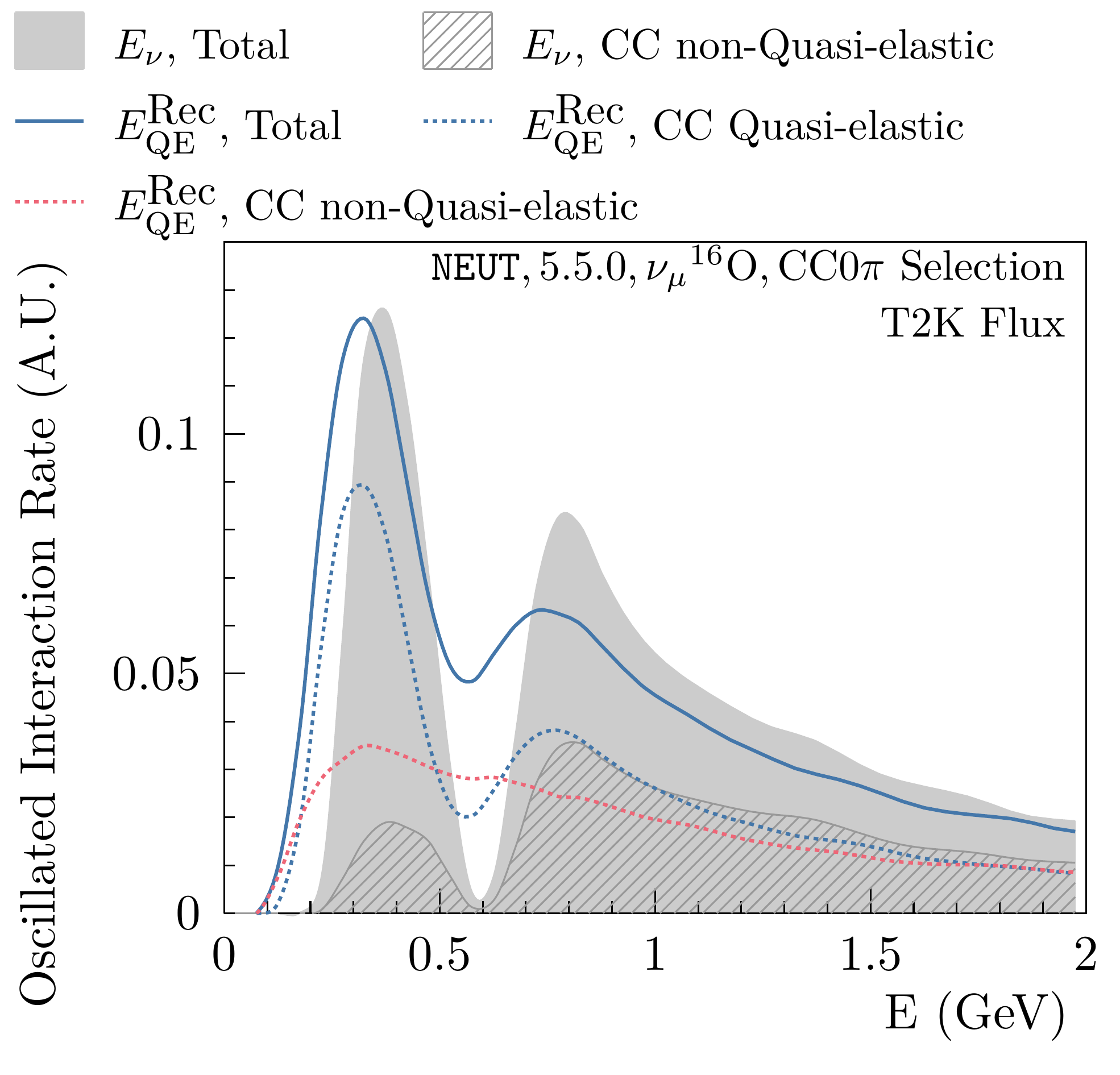}
  \includegraphics[width=0.8\textwidth]{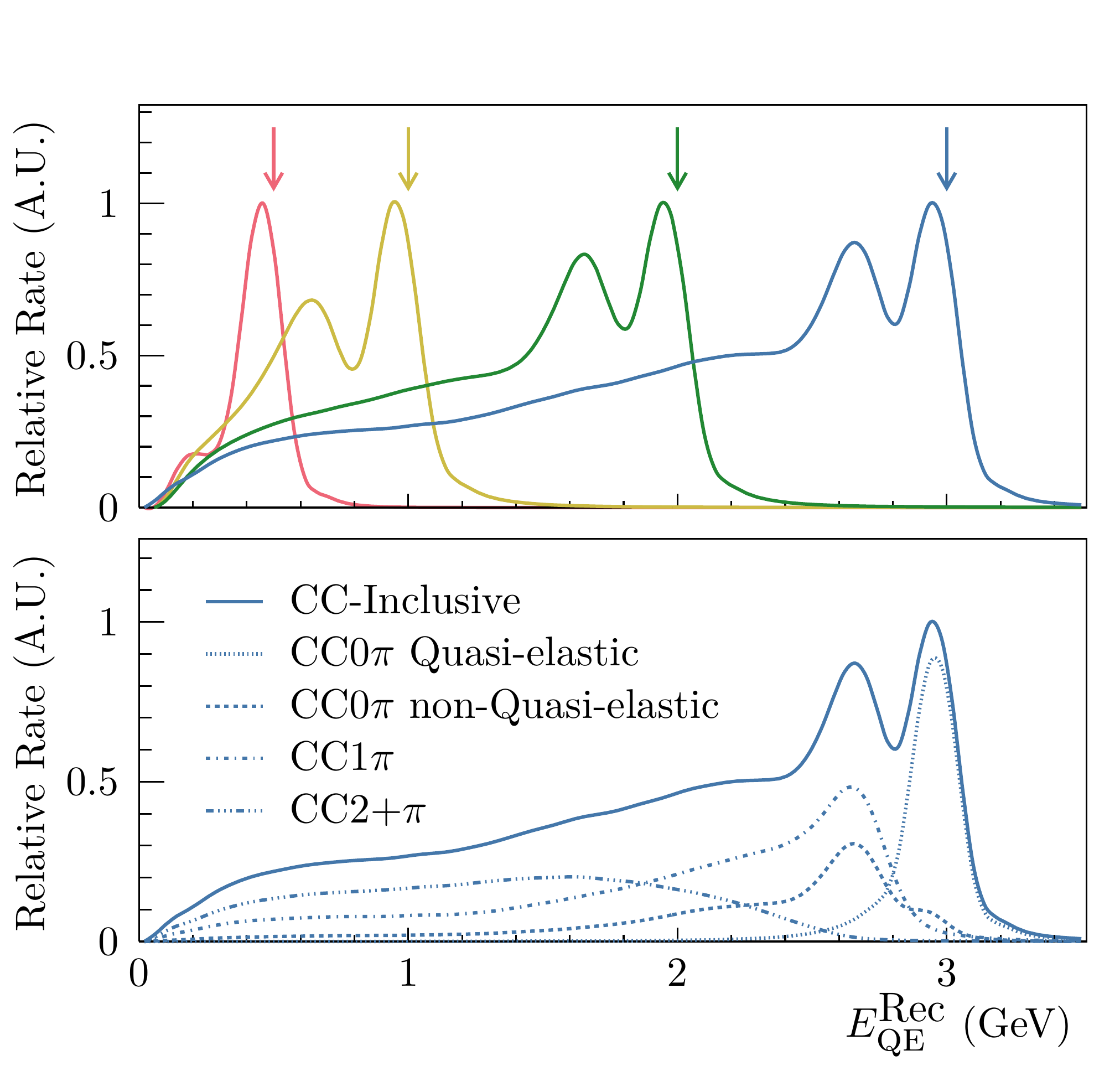}
  \caption{(top) The true and reconstructed neutrino energy for charged-current interactions producing no observable pion in a typical oscillated T2K $\nu_\mu$muon neutrino flux at \sk~\cite{T2K:2015sqm,T2K:2012nrr}. The reconstructed energy feed down for non-Quasi-elastic interactions smears away the oscillation signature that can be clearly seen in the true energy spectrum. (middle) The reconstructed energy feed down for four mono-energetic muon neutrino fluxes. The colored arrows show the true energy of the corresponding reconstructed energy distribution. (bottom) The 3~GeV distribution is separated by number of secondary mesons produced.}
  \label{fig:Introduction:Topo}
\end{figure}

To test oscillation hypotheses, observable distributions that correlate strongly with the neutrino energy distribution are sampled and compared to simulated predictions to extract oscillation parameter constraints. An example of such an observable is
\begin{equation}
\erecqe\left(p_{\ell},\theta_{\ell}\right)=\frac{2M_{N,i}E_{\ell} - M_{\ell}^{2} + M_{N,f}^{2} - M_{N,i}^{2}}{2\left(M_{N,i}-E_{\ell}+p_{\ell}\cos{\theta_{\ell}}\right)},
\end{equation}
where $M_{N,i}$, $M_{N,f}$, and $M_{\ell}$ are the mass of the initial-state nucleon, final-state nucleon, and final-state charged lepton respectively; $E_{\ell}$, $p_{\ell}$, and $\theta_{\ell}$ are the energy, three-momentum, and angle of the final-state charged-lepton respectively. \erecqe is an unbiased neutrino energy estimator for the charged-current quasi-elastic (CCQE) reaction off a free neutron, $\nu_\ell + n \rightarrow \ell^{-} + p$, or a free proton, 
$\bar{\nu_\ell} + p \rightarrow \ell^{+} + n$,
which is only a function of the kinematics of the final-state charged lepton. All operating neutrino-scattering experiments use nuclear targets, where CCQE reactions occur predominantly with bound nucleons. For such interactions, \erecqe is biased by the potential energy associated with the nucleon binding and is smeared by the Fermi motion of the nucleon confinement\footnote{This effect can be seen in the offset between the initial neutrino energy arrow and the peak of the \erecqe distribution in Fig.~\ref{fig:Introduction:Topo} (middle).}. For interactions that exhibit strong nuclear-effects or produce extra hadrons, here referred to as \emph{CC non-QE}, \erecqe is in general a significant underestimate of the true neutrino energy. Fig.~\ref{fig:Introduction:Topo} (top) shows the \neut prediction of the oscillated rate of charged-current neutrino--oxygen interactions that produce no observable pions, a CC0$\pi$ event \emph{topology}. The CCQE and CC non-QE components are shown separately to highlight how the shape of their contributions differ between the $E_\nu$ and \erecqe projections. For non-QE interactions, the shape of the oscillation is largely smeared away. The CC non-QE component of the CC0$\pi$ topology can be further broken down into interactions that either do not produce secondary pions, or do produce secondary pions that are subsequently re-absorbed before leaving the struck nucleus. Fig.~\ref{fig:Introduction:Topo} (middle and bottom) illustrates the \neut-predicted evolution of \erecqe for a relevant range of discrete neutrino energies. It can be seen that \erecqe exhibits significant reconstructed energy feed down above about 1~GeV, where secondary meson production is common. For higher energies, calorimetric estimators that account for hadronic energy are more accurate. However, current neutrino interaction models are often only predictive in lepton kinematics and significant uncertainties should be assigned to the predictions of hadronic particle spectra, especially for baryons and heavy mesons.

To accurately interpret observable distributions, interaction simulations are relied upon to predict the rate and observable projections for neutrino interactions over a range of energies and for a number of different target nuclei as well as the migration of these primary interactions into observable topologies. The next section details the nuclear dynamics, primary neutrino interaction, and hadronic re-scattering physics models implemented in \neut.

The most important proton decay channels, predicted by various
grand unification theories, produce a lepton
and a meson, \emph{e.g.} $p\rightarrow e^+ \pi^0$, $p\rightarrow \mu^+ \pi^0$, or $p\rightarrow \bar{\nu} K^+$. 
Neutrinos also often produce a lepton and one or more mesons when 
interacting with a nucleon. There are plenty of atmospheric 
neutrinos, which have sufficient energy to undergo 
such interactions, providing a significant background for proton decay searches. 
The majority of neutrino interactions 
also produce a baryon in addition to a lepton and a meson, 
but these may not be visible in detectors such as \sk.
The typical predicted momentum of mesons from both proton decays and 
atmospheric neutrino interactions are below several hundred MeV/c and thus, the probabilities for them to scatter before leaving the nucleus is high. Accurately predicting neutrino-induced backgrounds and hadronic re-scattering within the nucleus is critical for the sensitivity of proton decay searches.

\section{The Physics in \neut}\label{sec:Models}

\subsection{Simulating an Interaction}

\begin{figure}
  \centering
  \includegraphics[width=0.8\textwidth]{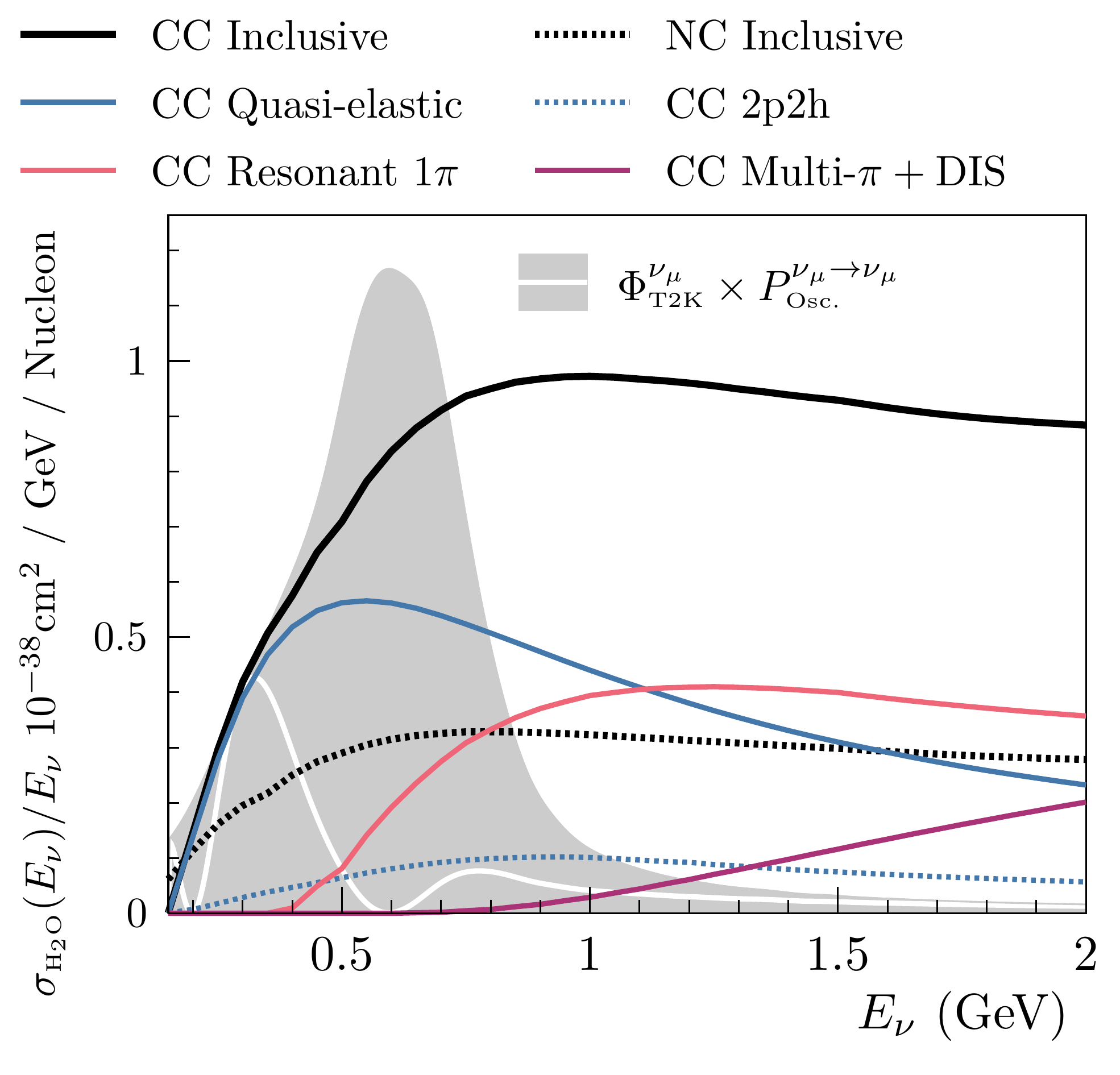}
  \includegraphics[width=0.8\textwidth]{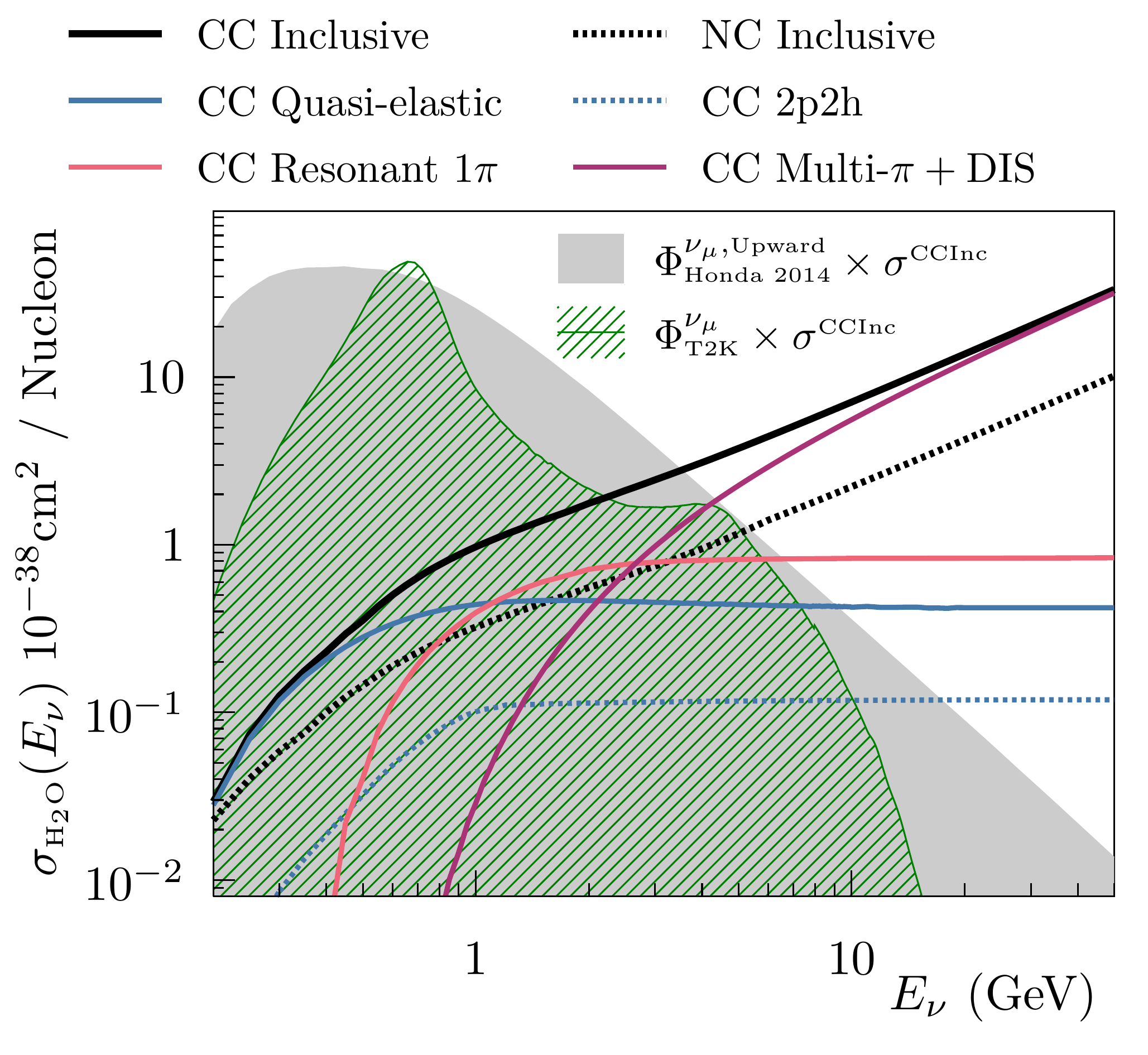}
  \caption{The \neut-predicted muon neutrino--water cross sections overlaid on the T2K muon neutrino flux~\cite{T2K:2012nrr}, with a typical oscillation (top), and the T2K and upward atmospheric muon neutrino fluxes~\cite{Honda:2015fha} multiplied by the charged-current inclusive total cross section (bottom). The flux multiplied by the cross section is proportional to the expected interaction rate. Above 4~GeV, the expected number of interactions in \sk arising from the T2K beam falls significantly faster than from atmospheric neutrinos.  \emph{n.b.} The cross sections presented in the top pane are divided by the neutrino energy, whereas in the bottom pane, they are not. This is to emphasise the saturation of the interaction channels associated with lower four-momentum transfer at \sk energies and the sharp turn-on seen over \ttwok flux distribution.
  }
  \label{fig:flux-xsec}
\end{figure}

In general, \neut factorizes the simulation of an interaction of a neutrino with flavour, $\ell$, and energy, $E_\nu$, into four discrete steps. First, a specific interaction channel is chosen randomly with probability, $P = \sigma_{T}^{i}\left(E_{\nu_\ell}\right)/\sigma_{T}^\textrm{tot}\left(E_{\nu_\ell}\right)$, where $\sigma_{T}^\textrm{tot}\left(E_{\nu_\ell}\right)$ is the total cross section and $\sigma_{T}^{i}\left(E_{\nu_\ell}\right)$ is the cross section for the specific target nuclei, $T$, and channel, $i$, where $i$ is an integer that identifies the interaction process and is defined in Table~\ref{tab:Models:channels_CC} (charged current) and Table~\ref{tab:Models:channels_NC} (neutral current). For neutrino--nucleon interaction channels, the nuclear-target cross section is usually constructed as $\sigma_{T}^{i} = Z\sigma_{p}^{i} + (A-Z)\sigma_{n}^{i}$, where $A$ and $Z$ are the nucleon number and the proton number of the target nuclei and $\sigma_{p}^{i}$ and $\sigma_{n}^{i}$ are the bound proton and bound neutron cross sections. For historical reasons, free protons can be added to nuclear targets to build simple molecular targets such as $\textrm{H}_2$O and CH. Fig.~\ref{fig:flux-xsec} shows the \neut water-target cross-section predictions separated into classes of interaction channel.

\begin{table}
\centering
\caption{Neutrino charged-current scattering reactions simulated by \neut, where p refers to a proton, n to a neutron, N to a neutron or proton, ${}^{A}\mathbf{X}$ to an entire nucleus, and $W$ to the invariant mass of the final state hadronic system.  The $\nu_\mu$ and $\bar{\nu_\mu}$ cross sections per nucleon for each process at 600 MeV and 10 GeV are included, where B.T. stands for Below Threshold and highlights kinematically disallowed channels for 600 MeV neutrinos. For reference, the \neut reaction identifier enumeration is included. CCQE (1p1h) cross-sections are for bound nucleon in oxygen and calculated using the spectral function model. Coherent pion production cross-sections are also for oxygen.}
\label{tab:Models:channels_CC}       
\begin{tabular}{ll|ll|c}
\hline\noalign{\smallskip}
& & \multicolumn{2}{c}{$\sigma^{\nu_{\mu}} / 10^{-38}\textrm{cm}^{2}$}/\textrm{N} & \\
Channel Name & Reaction & 0.6 GeV & 10 GeV & Enum \\
\noalign{\smallskip}\hline\noalign{\smallskip}
CCQE (1p1h) & $\nu_{\ell} + \textrm{n} \rightarrow \ell^{-} + \textrm{p} $ & 0.76 & 0.95 & \hphantom{-}1\hphantom{1} \\
     & $\bar{\nu}_{\ell} + \textrm{p} \rightarrow \ell^{+} + \textrm{n} $ & 0.20 & 0.85 & -1\hphantom{1} \\

2p2h & $\nu_{\ell} + \textrm{N}\textrm{n} \rightarrow \ell^{-} + \textrm{N}\textrm{p} $ & 0.03 & 0.08 & \hphantom{-}2\hphantom{1} \\
     & $\bar{\nu}_{\ell} + \textrm{N}\textrm{p} \rightarrow \ell^{+} + \textrm{N}\textrm{n} $ & 0.01 & 0.08 & -2\hphantom{1} \\

CCRes$1\pi^{+}$ & $\nu_{\ell} + \textrm{p} \rightarrow \ell^{-} + \textrm{p}\pi^{+} $ & 0.15 & 0.77 & \hphantom{-}11 \\
& $\nu_{\ell} + \textrm{n} \rightarrow \ell^{-} + \textrm{n}\pi^{+} $ & 0.03 & 0.52 & \hphantom{-}13 \\

CCRes$1\pi^{0}$ & $\nu_{\ell} + \textrm{n} \rightarrow \ell^{-} + \textrm{p}\pi^{0} $ & 0.04 & 0.39 & \hphantom{-}12 \\
& $\bar{\nu}_{\ell} + \textrm{p} \rightarrow \ell^{+} + \textrm{n}\pi^{0} $ & 0.01 & 0.31 & -12 \\

CCRes$1\pi^{-}$ & $\bar{\nu}_{\ell} + \textrm{n} \rightarrow \ell^{+} + \textrm{n}\pi^{-} $ & 0.02 & 0.63 & -11 \\
& $\bar{\nu}_{\ell} + \textrm{p} \rightarrow \ell^{+} + \textrm{p}\pi^{-} $ & $5\cdot{}10^{-3}$ & 0.41 & -13 \\

CCDif$1\pi^{+}$ & $\nu_{\ell} + \textrm{p} \rightarrow \ell^{-} + \textrm{p}\pi^{+} $ & B.T. & 0.03 & \hphantom{-}15 \\
CCDif$1\pi^{-}$ & $\bar{\nu}_{\ell} + \textrm{p} \rightarrow \ell^{+} + \textrm{p}\pi^{-} $ & B.T. & 0.03 & -15 \\

CCCoh$1\pi^{+}$ & $\nu_{\ell} + {}^{A}\mathbf{X} \rightarrow \ell^{-} + {}^{A}\mathbf{X} + \pi^+$ & $1\cdot{}10^{-3}$ & 0.04 & \hphantom{-}16 \\
CCCoh$1\pi^{-}$ & $\bar{\nu}_{\ell} + {}^{A}\mathbf{X} \rightarrow \ell^{+} + {}^{A}\mathbf{X} + \pi^{-} $ & $1\cdot{}10^{-3}$ & 0.04 & -16 \\

CCRes1$\gamma$ & $\nu_{\ell} + \textrm{n} \rightarrow \ell^{-} + \textrm{p}\gamma $ & $2\cdot{}10^{-4}$ & $2\cdot{}10^{-3}$ & \hphantom{-}17 \\
& $\bar{\nu}_{\ell} + \textrm{p} \rightarrow \ell^{+} + \textrm{n}\gamma $ & $3\cdot{}10^{-5}$ & $1\cdot{}10^{-3}$ & -17 \\

CCN$\pi$ & $\nu_{\ell} + \textrm{N} \rightarrow \ell^{-} + \textrm{N}^{\prime} + x\pi$ & B.T. & 0.85 & \hphantom{-}21 \\
& $\bar{\nu}_{\ell} + \textrm{N} \rightarrow \ell^{+} + \textrm{N}^{\prime} + x\pi$ & B.T. & 0.62 & -21 \\
& where $x > 1$ & & & \\
& and $1.3 < W < 2.0~\textrm{GeV}$ &&& \\

CCRes$1\eta^{0}$ & $\nu_{\ell} + \textrm{n} \rightarrow \ell^{-} + \textrm{p}\eta^{0} $ & B.T. & 0.19 & \hphantom{-}22 \\
& $\bar{\nu}_{\ell} + \textrm{p} \rightarrow \ell^{+} + \textrm{n}\eta^{0} $ & B.T. & 0.14 & -22 \\

CCRes$1K^{0}$ & $\nu_{\ell} + \textrm{n} \rightarrow \ell^{-} + \Lambda + K^{+} $ & B.T. & 0.06 & \hphantom{-}23 \\
CCRes$1K^{+}$ & $\bar{\nu}_{\ell} + \textrm{p} \rightarrow \ell^{+} + \Lambda + K^{0} $ & B.T. & 0.03 & -23 \\

CCDIS & $\nu_{\ell} + \textrm{N} \rightarrow \ell^{-} + \textrm{N}^{\prime} + x\pi$ & B.T. & 4.53 & \hphantom{-}26 \\
& $\bar{\nu}_{\ell} + \textrm{N} \rightarrow \ell^{+} + \textrm{N}^{\prime} + x\pi$ & B.T. & 1.30 & -26 \\
& where $W > 2.0~\textrm{GeV}$ && & \\

\noalign{\smallskip}\hline
\end{tabular}
\end{table}

\begin{table}
\centering
\caption{Neutrino neutral-current scattering reactions simulated by \neut, where p refers to a proton, n to a neutron, N to a neutron or proton, ${}^{A}\mathbf{X}$ to an entire nucleus, and $W$ to the invariant mass of the final state hadronic system. The $\nu_\mu$ and $\bar{\nu_\mu}$ cross sections per nucleon for each process at 600 MeV and 10 GeV are included, where B.T. stands for Below Threshold and highlights kinematically disallowed channels for 600 MeV neutrinos. For reference, the \neut reaction identifier enumeration is included. NCEL (1p1h) cross-sections are for bound nucleon in oxygen and calculated using the spectral function model. Coherent pion production cross-sections are also for oxygen.}
\label{tab:Models:channels_NC}       
\begin{tabular}{ll|ll|c}
\hline\noalign{\smallskip}
& & \multicolumn{2}{c}{$\sigma^{\nu_{\mu}} / 10^{-38}\textrm{cm}^{2}$}/\textrm{N} & \\
Channel Name & Reaction & 0.6 GeV & 10 GeV & Enum \\
\noalign{\smallskip}\hline\noalign{\smallskip}

NCRes$1\pi^{0}$ & $ \nu_{\ell} + \textrm{n} \rightarrow \nu_{\ell} + \textrm{n}\pi^{0}$ & 0.03 & 0.15 & \hphantom{-}31 \\
& & $9\cdot{}10^{-3}$ & 0.13 & -31 \\
& $ \nu_{\ell} + \textrm{p} \rightarrow \nu_{\ell} + \textrm{p}\pi^{0}$ & 0.03 & 0.15 & \hphantom{-}32 \\
& & $9\cdot{}10^{-3}$ & 0.12 & -32 \\

NCRes$1\pi^{0}$ & $ \nu_{\ell} + \textrm{n} \rightarrow \nu_{\ell} + \textrm{p}\pi^{-}$ & 0.02 & 0.13& \hphantom{-}33 \\
& & $5\cdot{}10^{-3}$ & 0.11 & -33 \\

NCRes$1\pi^{+}$ & $ \nu_{\ell} + \textrm{p} \rightarrow \nu_{\ell} + \textrm{n}\pi^{+}$ & 0.02 & 0.12 & \hphantom{-}34 \\
& & $5\cdot{}10^{-3}$ & 0.10 & -34 \\

NCDif$1\pi^{0}$ & $\nu_{\ell} + \textrm{p} \rightarrow \nu_{\ell} + \textrm{p}\pi^{0} $ & B.T. & 0.01 & \hphantom{-}35 \\
& & B.T. & 0.01 & -35 \\

NCCoh$1\pi^{0}$ & $\nu_{\ell} + {}^{A}\mathbf{X} \rightarrow \nu_{\ell} + {}^{A}\mathbf{X} + \pi^{+} $ & $1\cdot{}10^{-3}$ & 0.02 & \hphantom{-}36 \\
& & $1\cdot{}10^{-3}$ & 0.02 & -36 \\

NCRes1$\gamma$ & $\nu_{\ell} + \textrm{n} \rightarrow \nu_{\ell} + \textrm{n}\gamma $ & $2\cdot{}10^{-4}$ & $1\cdot{}10^{-3}$ & \hphantom{-}38 \\
& & $6\cdot{}10^{-5}$ & $8\cdot{}10^{-4}$ & -38 \\

& $\nu_{\ell} + \textrm{p} \rightarrow \nu_{\ell} + \textrm{p}\gamma $ & $2\cdot{}10^{-4}$ & $9\cdot{}10^{-4}$ & \hphantom{-}39 \\
& & $6\cdot{}10^{-5}$ & $8\cdot{}10^{-4}$ & -39 \\

NCN$\pi$ & $\nu_{\ell} + \textrm{N} \rightarrow \nu_{\ell} + \textrm{N}^{\prime} + x\pi$ & $2\cdot{}10^{-4}$ & 0.25 & \hphantom{-}41 \\
& & $3\cdot{}10^{-5}$ & 0.23 & -41 \\
& where $x > 1$ &  &&\\
& and $1.3 < W < 2.0~\textrm{GeV}$ & && \\

NCRes1$\eta^{0}$ & $\nu_{\ell} + \textrm{n} \rightarrow \nu_{\ell} + \textrm{n}\eta^{0} $ & B.T. & 0.04 & \hphantom{-}42 \\
& & B.T. & 0.03 & -42 \\

& $\nu_{\ell} + \textrm{p} \rightarrow \nu_{\ell} + \textrm{p}\eta^{0} $ & B.T. & 0.03 & \hphantom{-}43 \\
& & B.T. & 0.02 & -43 \\

NCRes1$K^{0}$ & $\nu_{\ell} + \textrm{n} \rightarrow \nu_{\ell} + \Lambda + K^{0} $ & B.T. & 0.01 & \hphantom{-}44 \\
& & B.T. & $7\cdot{}10^{-3}$ & -44 \\
NCRes1$K^{+}$ & $\nu_{\ell} + \textrm{p} \rightarrow \nu_{\ell} + \Lambda + K^{+} $ & B.T. & $9\cdot{}10^{-3}$ & \hphantom{-}45 \\
& & B.T. & $7\cdot{}10^{-3}$ & -45 \\

NCDIS & $\nu_{\ell} + \textrm{N} \rightarrow \nu_{\ell} + \textrm{N}^{\prime} + x\pi$ & B.T. & 1.37 & \hphantom{-}46 \\
& & B.T. & 0.49 & -45 \\
& where $W > 2.0~\textrm{GeV}$ & && \\

NCEL (1p1h) & $\nu_{\ell} + \textrm{n} \rightarrow \nu_{\ell} + \textrm{n} $ & 0.13 & 0.16 & \hphantom{-}51 \\
& & 0.05 & 0.16 & -51 \\

& $\nu_{\ell} + \textrm{p} \rightarrow \nu_{\ell} + \textrm{p} $ & 0.16 & 0.21 & \hphantom{-}52 \\
& & 0.06 & 0.19 & -52 \\

\noalign{\smallskip}\hline
\end{tabular}
\end{table}

Second, the primary neutrino interaction, or \emph{hard scatter}, is simulated. For the majority of channels, this step involves choosing a bound nucleon from an initial-state nuclear model, then choosing interaction kinematics according to the specific interaction model, and finally choosing any remaining particle kinematics not specified by the model. This step is performed under the impulse approximation~\cite{PhysRevD.72.053005}, which treats the target bound nucleon and the remnant nucleus as evolving independently during and after the hard scatter. This further factorizes the simulation as, to first order, the sampling of the nuclear model does not depend on the interaction kinematics chosen.

For the coherent pion-production channels (Enum 16 and 36), the interaction occurs coherently between the neutrino and the target nucleus and as a result no bound nucleon target is chosen and this is considered the final step of the simulation. For other channels, the final state hadrons are then passed on to the third step, the nucleon and meson intra-nuclear re-scattering simulation, where hadrons can elastically scatter, exchange charge with a nucleon in the nucleus, or be produced or absorbed as they are stepped out of the nuclear medium. 

Finally, for oxygen targets only, the final state nuclear remnant can be left in an excited state after the interaction and a number of nuclear de-excitations, producing low energy photons ($\mathcal{O}\left(1-10\right)\,\textrm{MeV}$), are modeled following Ref.~\cite{PhysRevC.48.1442}. Careful treatment of the de-excitation oxygen is important for precisely simulating interactions in the sensitive \sk detector.

For the majority of particles produced in the hard scatter and subsequent re-scattering, \neut stores their properties in an \textit{event vector} file that can be used as input to further experiment simulation processes. The only exceptions are tau and omega particles, which are decayed during the \neut simulation by TAUOLA\cite{Jadach:1993hs} and a custom omega decay simulation code, respectively. The final states of such decays are written to the event vector.

\subsection{The Hard Scatter}

This section details the neutrino scattering physics models implemented in \neut, roughly ordered by the degree of inelasticity of the interaction.

\subsubsection{Quasi-Elastic Scattering}\label{sec:Models:CCQE}

The cross-section of neutrino--nucleon charged-current 
quasi-elastic (CCQE) interaction was formalized 
by Llewellyn Smith~\cite{QECRSFN}. In some of the literature, an equivalent channel is referred to as 1p1h, for one-particle one-hole.
Three different nuclear models are implemented in \neut for simulating CCQE interactions, the relativistic global Fermi-gas (GRFG) model, 
the local Fermi-gas (LFG) model and the spectral function (SF) model. 

The \neut implementation of the GRFG coss-section follows the prescription by Smith and Moniz~\cite{QECRSFRMGAS}.  The LFG implementation uses the model by Nieves~\emph{et al.}~\cite{Gran:2013kda} and includes an updated removal energy treatment implemented by Bourguille~\emph{et al.}~\cite{Bourguille:2021:04}.
Simple Fermi-gas models tend to over-predict the cross section for forward going leptons, as a result this model takes into account long and short range correlation of nucleons using the random phase approximation, suppressing the cross-section for low 4-momentum transfer~\cite{Nieves:2004wx}.
The \neut SF uses the spectral function by Benhar~\emph{et al.}~\cite{Benhar:1994hw} 
and the implementation is based on the one in NuWro~\cite{NuWro2006} with additional 
improvements by Furmanski~\cite{Furmanski:2015knr}. 

\begin{figure}
  \centering
  \includegraphics[width=0.8\textwidth]{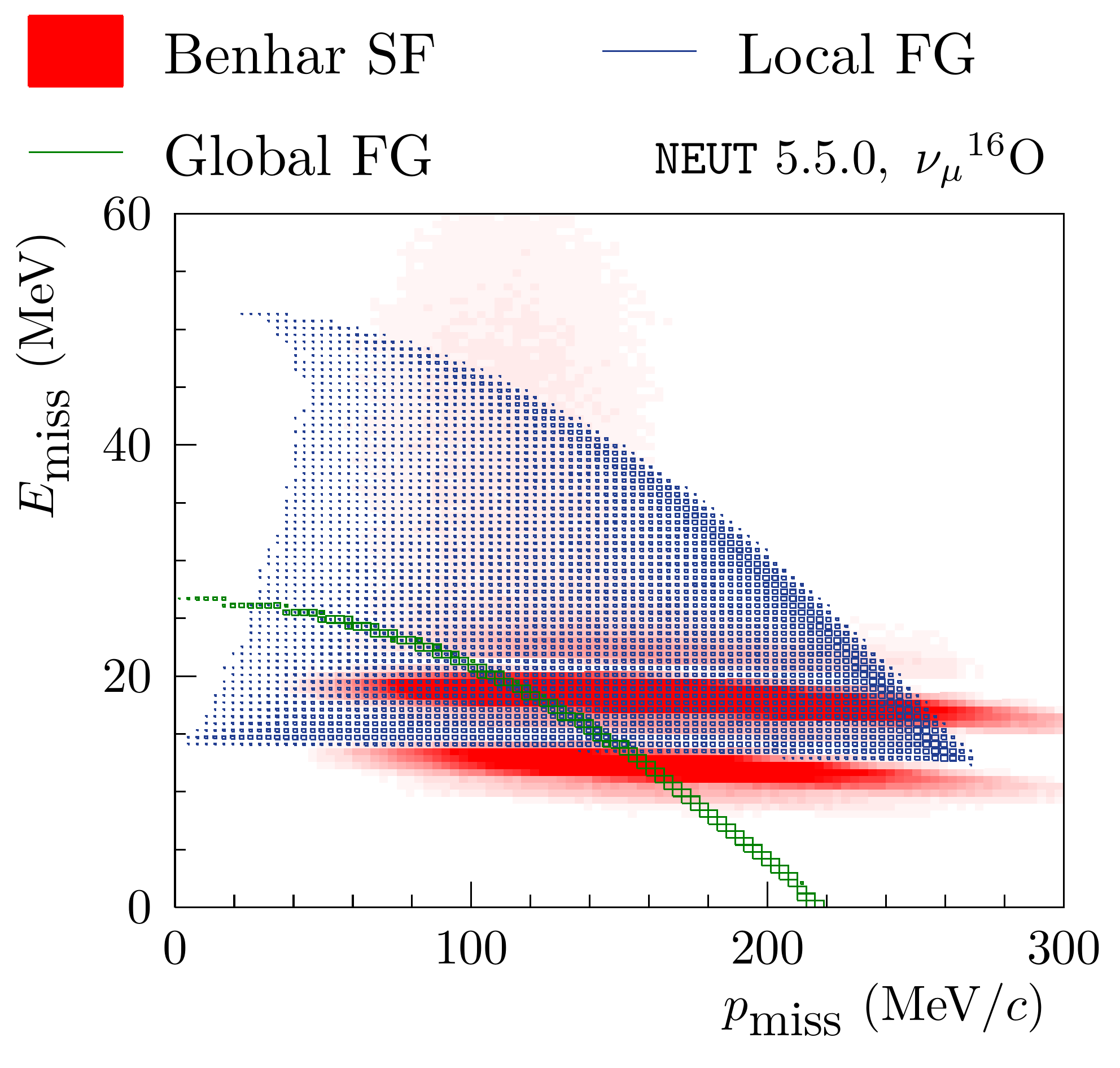}
  \includegraphics[width=0.8\textwidth]{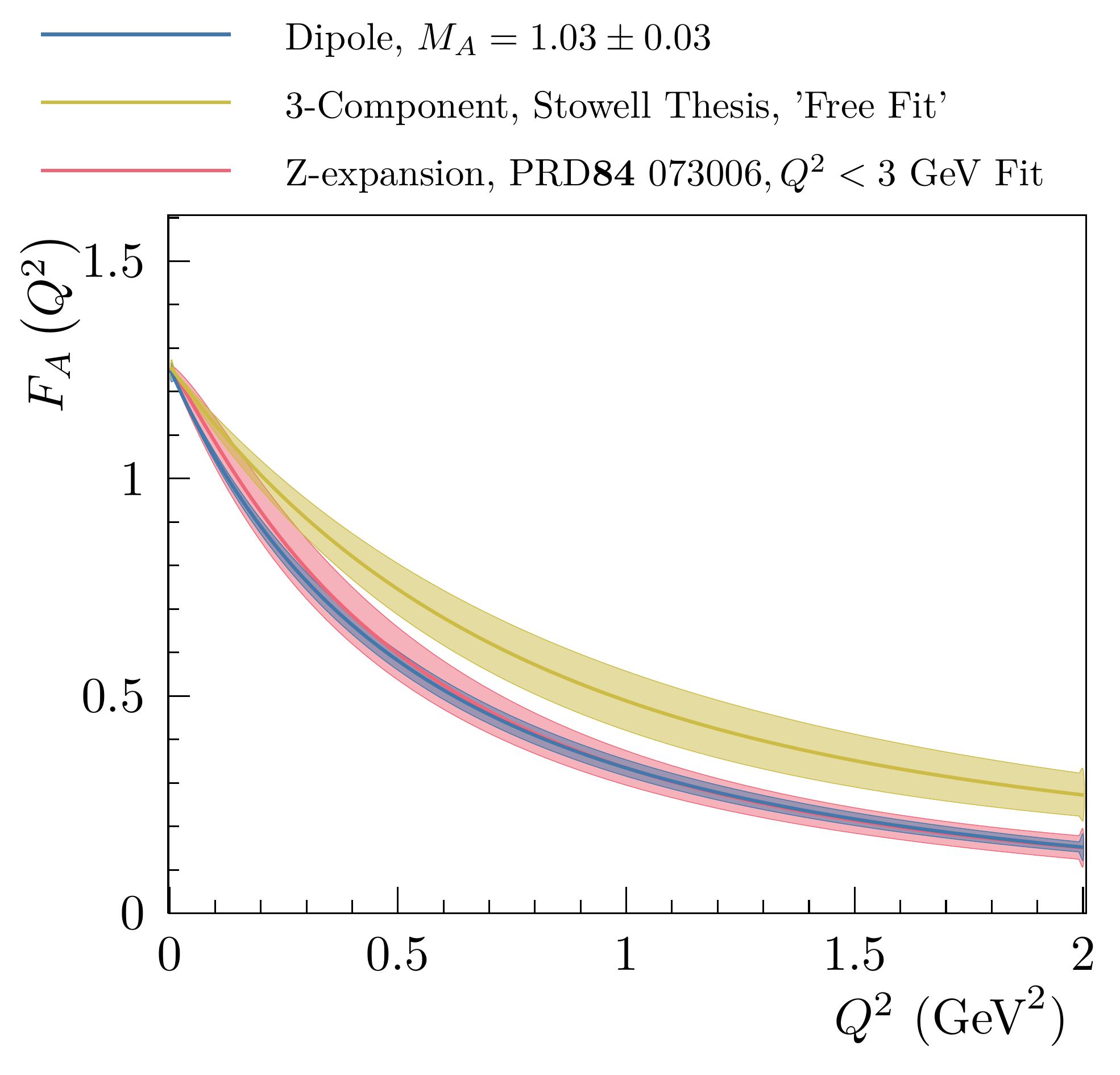}
  \caption{(Top) The reconstructed missing energy and missing momentum distributions for the three initial-state nuclear models implemented for the CCQE channel. (Bottom) Three nucleon axial form factor models and their associated uncertainties derived in Ref.~\cite{StowellThesis} (Dipole and 3-component) and Ref.~\cite{PhysRevD.93.113015} (Z-expansion). }
  \label{fig:nuclear-model}
\end{figure}

These three nuclear models differ in their treatment of the bound nucleon momentum and removal energy 
distributions and whether there are correlations between them. Fig.~\ref{fig:nuclear-model} 
shows the projections of interactions simulated with each of the three models into missing
momentum ($p_\textrm{miss}$) and missing energy ($E_\textrm{miss}$), which are observable 
quantities for analogous measurements of electron--nuclear scattering. 
For the $\nu_\ell + {}^{16}\textrm{O} \rightarrow \ell^{-} + p + {}^{15}\textrm{O}$ interaction, 
$p_\textrm{miss}$ and $E_\textrm{miss}$ are defined as

\begin{align}
  p_\textrm{miss} &= \left|\vec{p}_\nu - \vec{p}_\ell - \vec{p}_p\right|, \\
  T_{{}^{15}\textrm{O}} &= \sqrt{p_\textrm{miss}^{2} + M_{{}^{15}\textrm{O}}^{2}} - M_{{}^{15}\textrm{O}},\\  
  E_\textrm{miss} &= E_\nu + M_n - E_\ell - E_p - T_{{}^{15}\textrm{O}},
\end{align}

where $T_{{}^{15}\textrm{O}}$ is the reconstructed kinetic energy and $M_{{}^{15}\textrm{O}}$ the ground-state mass of the nuclear remnant, which in this case is an unstable isotope but is long-lived on the timescale of the impulse approximation. These quantities are of interest because $E_\textrm{miss}$ is approximately the energy lost to the nuclear response during the interaction and $p_\textrm{miss}$ is the momentum of the struck nucleon in the lab frame. It is clear from Fig.~\ref{fig:nuclear-model} that the different nuclear models make different predictions about their distribution and correlation, notably the ${}^{16}\textrm{O}$ shell structure is visible in the SF predictions. The SF model is tuned to exclusive electron--nuclear scattering data and is only available for a subset of target nuclei, and in \neut is only implemented for ${}^{12}\textrm{C}$, ${}^{16}\textrm{O}$, and ${}^{56}\textrm{Fe}$, the most important nuclei for \sk and \ttwok analyses. For interactions with other nuclei, such as ${}^{40}\textrm{Ar}$, only the GRFG or the LFG nuclear models are implemented.

Beyond the choice of nuclear models, the vector and axial-vector nucleon form factors control the strength and shape of quasi-elastic interactions.
There are two vector form factors of nucleon implemented in \neut,
the simple dipole form factor and BBBA05~\cite{Bradford:2006yz}. 
By default, BBBA05 is used as this form factor was developed to reproduce experimental electron-scattering data. 

There are four axial nucleon form factor models implemented, the dipole form factor, BBBA07~\cite{Bodek:2007ym},
the Z-expansion model~\cite{Bhattacharya:2011} and 
the 3-component model~\cite{StowellThesis}. The 3-component fit model is inspired
by the 2-component fit model~\cite{Adamuscin:2007fk} and was created to provide
additional shape freedom by expanding the 2-component model,
which quickly decays with four-momentum transfer squared, $Q^{2}$. 
This model has the freedom to vary the 
gradient of the form factor at low-$Q^{2}$ leaving the free parameters
in the 2-component model to set the shape at higher momentum transfer.
The 3-component model is able to be continuously varied between
the shape of both the 2-component and the simple dipole model. Fig.~\ref{fig:nuclear-model} presents the shape of three of the axial form factor models with associated uncertainties derived fits to hydrogen and deuterium bubble chamber data. The dipole and 3-component model uncertainties are reproduced from Stowell~\cite{StowellThesis}, and the Z-expansion model was fit by Meyer~\emph{et. al.}~\cite{PhysRevD.93.113015}.

For neutral current elastic (NCEL) scattering, the treatment and available model components are equivalent to CCQE, except that an LFG initial-state model is not implemented for NCEL.

\subsubsection{Charged Current Multi-Nucleon Scattering}\label{sec:Models:CCMultiNucl}

A number of accelerator-based neutrino oscillation experiments, K2K, MINOS, and MiniBooNE, started taking high-statistics neutrino--nuclear scattering data in the 2000s. These experiments found that
the number of the observed CCQE-like (equivalent to the CC0$\pi$ topology introduced earlier) events were a few tens of percent
larger than predicted by the models, but with a relative deficit of very forward-going muons~\cite{K2K:2006odf,MiniBooNE:2013qnd,MINOS:2014axb}.
One of the sources of these 
discrepancies was thought to be coming from neutrino--nucleus interaction channels, 
which were not implemented in the simulations used. The most-probable candidate is now believe to be
the so-called \emph{multi-nucleon} interaction, of which a similar process
is known to exist in electron--nucleus scattering. Inclusion of this
interaction into neutrino--nucleus simulations was discussed by Marteau in 1999~\cite{Marteau:1999kt}.

In \neut, the Valencia model by Nieves~\emph{et al.}~\cite{Gran:2013kda} is implemented. This
model considers an interaction involving the production of two nucleons and two
holes in a ground-state nuclear target (often called a 2p2h interaction in the literature). 
Their model includes processes involving the exchange of mesons between
two nucleons and thus, sometimes, it is often referred to as a Meson Exchange Current, or MEC, model.
The model is not applicable for large momentum 
transfer and thus the three-momentum transfer to the nucleus is limited to
$q_3 < 1.2$~GeV/c. This model does not predict how the four-momentum is distributed between the two final-state nucleons, \neut follows the implementation in NuWro~\cite{NuWro2006}.
The directions of the outgoing nucleons are selected to be
uniformly distributed in the center of mass frame of the nucleons.
A separation energy (sometimes ambiguously referred to as the \emph{binding energy}) is subtracted from the energy transfer from the lepton system. The outgoing nucleon momenta are required to be larger than the local Fermi surface at the interaction position within the nucleus. The model for the binding energy is described in Ref.~\cite{Bourguille:2021:04} and depends on the interaction position in the nucleus. For oxygen, typically between $\sim$50~MeV and $\sim$75~MeV of energy is lost to the nuclear response.

Neutral current multi-nucleon scattering is not yet implemented in
\neut.

\subsubsection{Single Meson and Gamma Productions}\label{sec:Models:1Meson}

Single pion production
is one of the dominant neutrino interaction channels
in the few-GeV energy region. Therefore, it is important to 
understand these interactions to study neutrino oscillations
using the atmospheric or the accelerator neutrinos.
The particles from the single
meson productions are similar to the ones from nucleon decay, 
and thus, these interactions are also important in the 
nucleon decay searches.

Single pion production in \neut is implemented following the 
model by Rein and Sehgal~\cite{Rein:1980wg}.
An improved model, which takes into account the lepton mass
correction, by Berger and Sehgal~\cite{Berger:2007rq} is 
also implemented. Both of the models simulate these
interactions in two steps. First, a neutrino
excites the nucleon and produce intermediate baryon resonance state, which then decays into a single
meson or gamma and baryon. The range of the invariant mass of the 
intermediate state, $W$, is between the pion-production
threshold and smaller than 2 $\rm{GeV/c^2}$.
In these two models, two kinds of form factors are implemented. 
The first is a simple dipole form,
as used in the original publications by Rein and 
Sehgal~\cite{Rein:1980wg}. The axial vector mass of the dipole 
type form factor was selected to be $1.21 \rm{GeV/c^2}$ based 
on fits to K2K data.
The second form was formalized by Graczyk and 
Sobczyk~\cite{Graczyk:2007xk} based on the Rarita-Schwinger 
formalization. The free parameters of Graczyk-Sobczyk form 
were extracted by fits to hydrogen and deuterium bubble chamber data.
The direction of the pions in the resonance rest frame, 
the so-called Adler frame, can be determined using the full prescription 
by Rein~\cite{Rein:1987cb} for all implemented resonances. Alternatively, the Rein calculation can be only used for interactions producing an intermediate $\Delta(1232)$ while, for higher-order resonances,
final-state pions are distributed uniformly in the Adler frame.
For nuclear-target interactions, Pauli-blocking is considered and final-state nucleon produced in the resonance decay is required to have the momentum larger than the local Fermi surface
momentum modelled by an LFG. The overall effect is
small, typically less than a few percent of the events are rejected.

Single Kaon, Eta and $\gamma$ productions are simulated using 
the same framework for the single pion production. The main
differences are the decay probabilities (branching ratios) of each simulated resonance to the
relevant final state. The production of other mesons, such as the omega, is not simulated with the model described in this section. However, such particles can be produced during the hadronization simulation in the Deep Inelastic Scattering reactions, described in the next section.

For neutrino interactions with nucleon bound within a nucleus,
produced hadrons undergo the final state interactions as described
in Sec.~\ref{sec:HadronInteraction}. Additionally,
formation zone effects are taken into account as described in
Sec.~\ref{sec:Models:FormationZone}.

\subsubsection{Shallow and Deep Inelastic Scattering}\label{sec:Models:SISDIS}

Shallow and deep inelastic scattering processes are separated
as shown in Table~\ref{SISDIS:ModeTable} to avoid double counting single- and multi-meson production channels.

\begin{figure}
  \centering
  \includegraphics[width=1\textwidth]{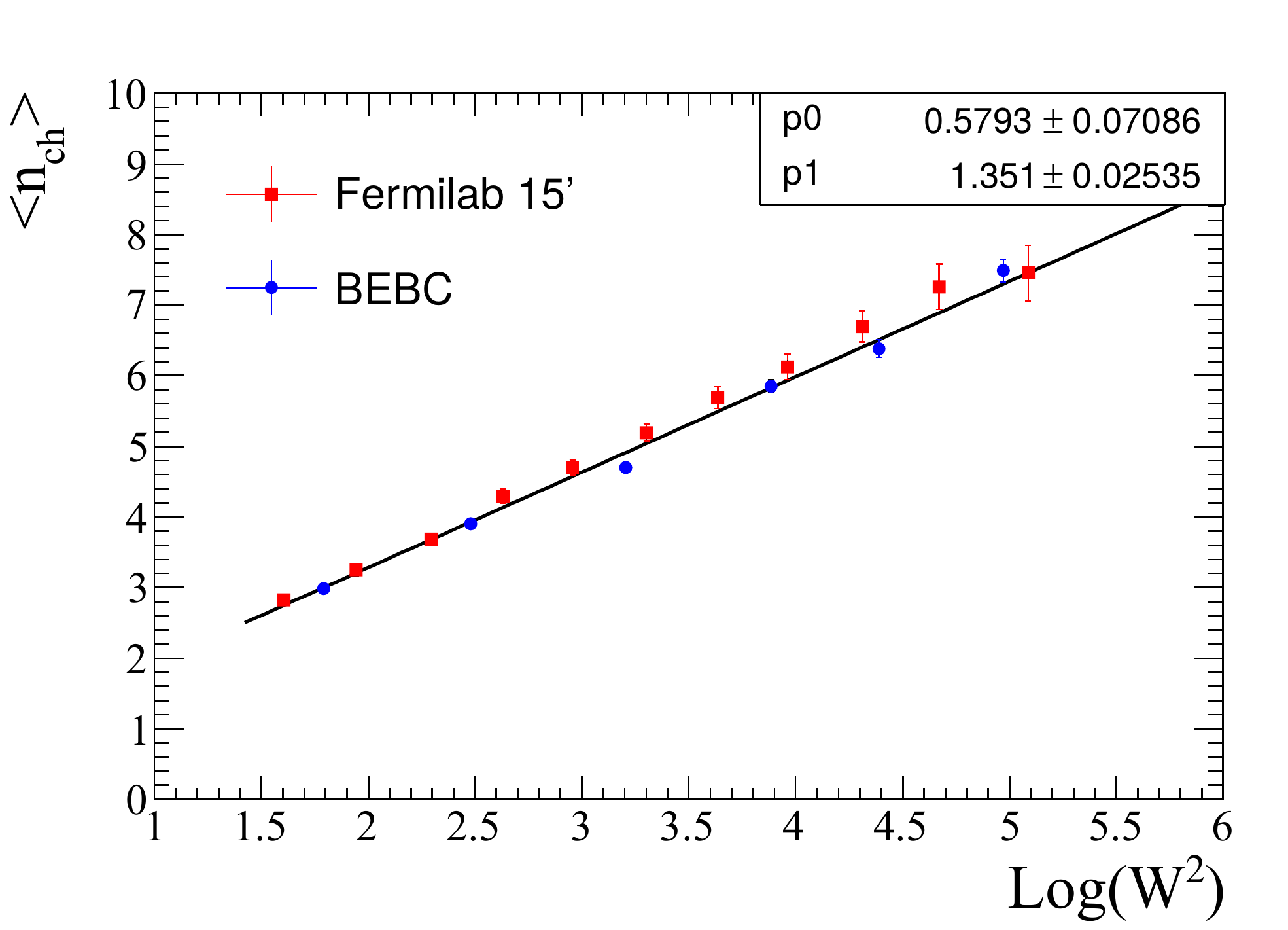}
  \caption{Results of the fit of the averaged charged hadron multiplicities
  as a function of $W$ using the bubble chamber data sets~\cite{MPIMULTP,MPIMULTP_BEBC}.
  Figure reproduced from Ref.~\cite{Bronner:2016gmz}.}
  \label{fig:MPiDIS-multiplicity-models}
\end{figure}

\begin{table}
\caption{Shallow and deep inelastic scattering implementation in \neut. 
$W$ is the invariant mass of the intermediate hadron system.}
\label{SISDIS:ModeTable}
\centering
\begin{tabular}{l|l|l}
\hline
  & 1 meson  & more than 1 meson  \\
\hline
$W< 2\rm{GeV/c^2}$  & (covered by single & custom multi-pion\\ 
  & meson production ) & production model \\
\hline
$W >2\rm{GeV/c^2}$  & \multicolumn{2}{c}{Pythia v5.72, included in \cernlib 2005} \\
\hline
\end{tabular}
\end{table}
Interactions producing more than one meson is simulated
with a custom  multi-pion-production model. When $W$ is larger than 2~$\rm{GeV/c^2}$,
all the interactions, which produced at least one meson are simulated
by PYTHIA v5.72~\cite{Sjostrand:1993yb}, included in \cernlib 2005.
For both cases, same parton distribution functions (PDFs) are used.
The default PDF is a modified version of GRV98~\cite{Gluck:1998xa} that is
based on the Bodek and Yang model~\cite{Bodek:2005de}.
The multi-pion production cross section is a function of the pion multiplicity,
which itself is a function of $W$, that models the probability to
produce more than one pion, with $W < 2$~GeV/$c^{2}$ region. The enforcement of multiple final-state pions avoids overlap with the resonance single meson production channels.
The mean multiplicity of charged hadrons 
is estimated from the results of Fermilab 15-foot hydrogen bubble chamber
experiment~\cite{MPIMULTP} 
and the Big European Bubble Chamber (BEBC) experiment~\cite{MPIMULTP_BEBC}.
Two parameter sets are available. The first uses the data
from~\cite{MPIMULTP} and predicts
\begin{equation}
  \langle \textrm{n}_\textrm{ch}\rangle=0.09 + 1.83\times\mbox{ln}(W^2),
  \label{eqn:MEANMULTIP1}
\end{equation}
for all channels.
The second uses the fit results from both~\cite{MPIMULTP}
and~\cite{MPIMULTP_BEBC}. This time, the parameter sets obtained by Bronner~\cite{Bronner:2016gmz} treat each combination of $\nu+p$, $\nu+n$, $\bar{\nu}+p$ and $\bar{\nu}+n$ separately
by fitting the measured average charged hadron multiplicity data for each channel independently.
The Bronner model for the $\nu+p$ channel is reproduced in Fig. \ref{fig:MPiDIS-multiplicity-models}.
The obtained parameter sets are 
\begin{eqnarray}
  \langle \textrm{n}_\textrm{ch}\rangle&=&0.58 + 1.35\times\mbox{ln}(W^2) ~\textrm{for}~\nu+p, \\
  \langle \textrm{n}_\textrm{ch}\rangle&=&0.35 + 1.24\times\mbox{ln}(W^2) ~\textrm{for}~\nu+n, \\
  \langle \textrm{n}_\textrm{ch}\rangle&=&0.41 + 1.18\times\mbox{ln}(W^2) ~\textrm{for}~\bar{\nu}+p, \\
  \langle \textrm{n}_\textrm{ch}\rangle&=&0.80 + 0.94\times\mbox{ln}(W^2) ~\textrm{for}~\bar{\nu}+n.
  \label{eqn:MEANMULTIP2}
\end{eqnarray}

For such interaction, we assume KNO scaling to determine the value of 
$W$~\cite{MPIWDIST}. The forward-backward asymmetry of the charged hadron multiplicity 
in the hadronic center of mass system is modelled as
\begin{equation}
  \frac{\textrm{n}_\textrm{ch}^F}{\textrm{n}_\textrm{ch}^B}=
  \frac{0.35+0.41\times\mbox{ln}(W^2)}{0.5+0.09\times\mbox{ln}(W^2)},
\end{equation}
which was derived from BEBC data~\cite{MPIFBASYM}.

The cross-section for the events with $W$ larger than 2~$\rm{GeV/c^2}$
is calculated without the multiplicity factor because there is no
overlap with other implemented models. The kinematics of the produced particles are determined by PYTHIA.

For neutrino interactions with a nucleon bound within a nucleus,
final state hadron interactions and formation zone effects are taken into
account as for single meson production.

\subsubsection{Coherent and Diffractive Pion Productions}\label{sec:Models:COHDIF}
There are two coherent neutrino--nucleus pion production models implemented in \neut. 
The default model is based on the prescription by 
Berger and Sehgal~\cite{Berger:2008xs}. This is an update to the previously publication model by Rein and Sehgal~\cite{Rein:1982pf},
which is also implemented in \neut. The new model uses improved
elastic pion--carbon cross-section data and lepton-mass effects are properly
taken into account. With these improvements, the model is applicable
to neutrinos with energies below several GeV.

A similar process, diffractive pion
production, involves coherent-like pion production, but with a single proton.
\neut implements the prescription by Rein~\cite{Rein:1986cd}.

Pions produced through these channels are not affected by the final state
re-scattering, which is described in the following section \ref{sec:HadronInteraction:meson}.

\subsubsection{Formation zone}\label{sec:Models:FormationZone}

The idea of a \emph{formation zone} is implemented and the production positions of hadrons in
nucleus are shifted from the initial neutrino interaction 
position. The implemented model in \neut is based on 
SKAT data~\cite{Baranov:1985mb,Ammosov:2001}.
The production points of the hadrons for those interactions are 
shifted using the formation length ($L_{FZ}$), where
\begin{equation}
    L_{FZ}=p/\mu^2,
\end{equation}
$p$ is the momentum of the hadron and $\mu=0.08$(GeV/c$^2$). 
The actual size of the shift is determined as 
$L_{FZ}\times(-\log({\rm rand}[0,1])$, where rand[0,1] is 
a random number from 0 to 1. The distribution of secondary hadron production position is shifted further from the center of the nucleus,  
the produced position of hadrons shifts to the outer region 
of the nucleus and thus reduce the probability of FSI.

\subsection{Final-state Hadronic Re-scattering}\label{sec:HadronInteraction}

\neut simulates the interactions of pions, kaons, etas, omegas, protons and
neutrons, produced via neutrino interactions
or nucleon decay, within the nucleus. In order to simulate these hadron 
interactions, a custom semi-classical intranuclear cascade (INC) 
model is used, as in most other neutrino--nucleus simulations. In \neut, a hadron produced in the nucleus is 
tracked step by step from the production point until the particle
escapes from the nucleus. The size of each step is fixed at 0.2~fm.
At each step, it is decided whether the particle has interacted
or not using the mean free paths for the modelled interaction channels.
A Woods-Saxon nucleon density function and the local 
Fermi-gas model are used to determine the interaction positions
and kinematics of the initial and final states.

\subsubsection{Meson interactions in nucleus}\label{sec:HadronInteraction:meson}

Among the modelled mesons, pion is the most important in the analyses
of SK and T2K.
The mean free paths for the pion interactions in the Delta
region are calculated following the prescriptions by 
Salcedo~\emph{et al.}~\cite{Salcedo:1987md}. 
Their model takes into account the in-medium correction of
$\Delta$ self energy and uses the local Fermi-gas model.
Therefore, the obtained mean free paths are expressed
as functions of nuclear density and pion momentum.
This model is applicable for the pions with momenta smaller
than 500~MeV/c.
The mean free paths for pions with  momenta larger 500~MeV/c
was motivated by fits to pion--nucleon scattering data and are expressed
as functions of pion momentum only.

\begin{figure}
  \centering
  \includegraphics[width=1\textwidth]{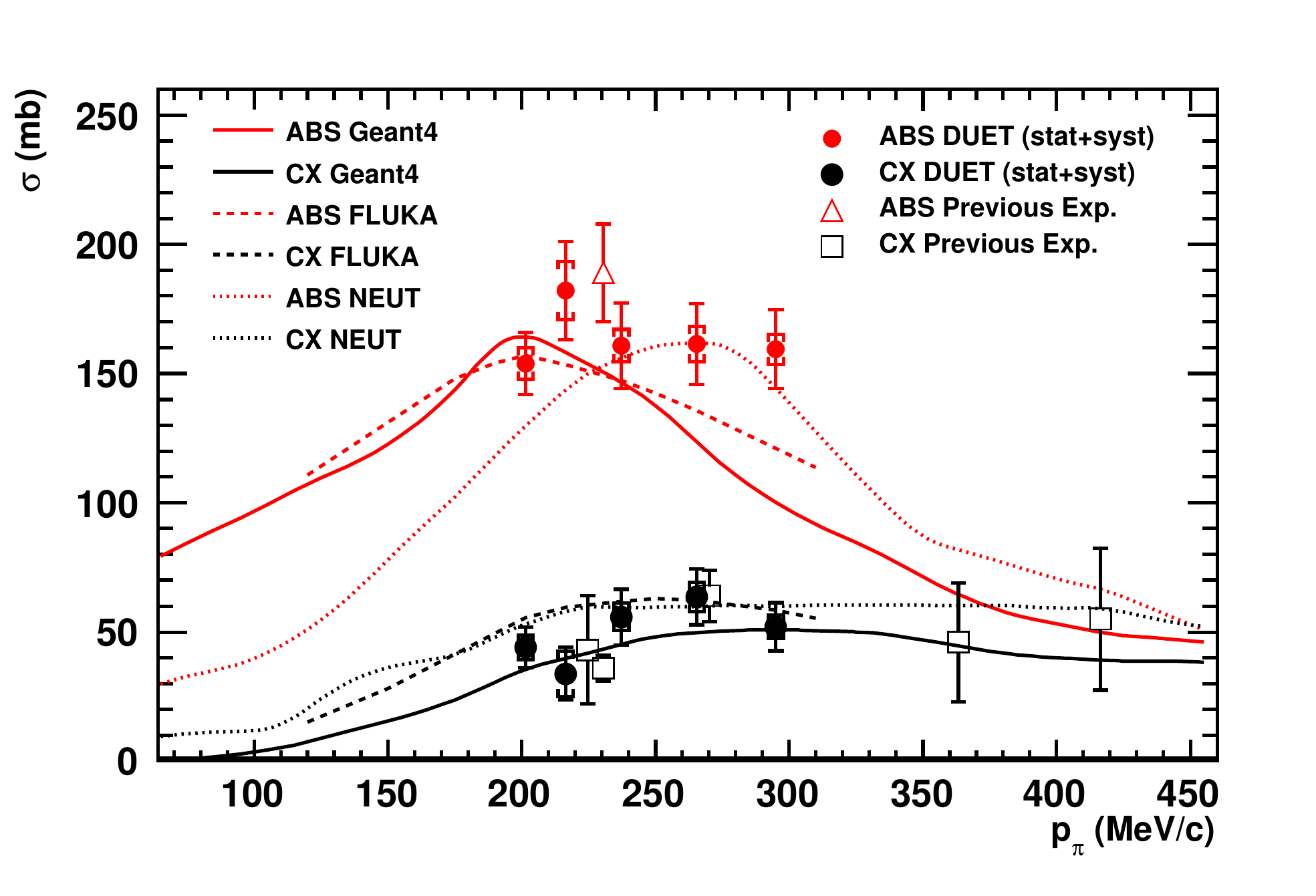}
  \caption{Comparisons of the $\pi^+$ absorption and charge exchange in Carbon cross-sections data and the simulated results of three simulation programs, GEANT4 (solid line), FLUKA (dashed line), and \neut (dotted line). The data are taken from~\cite{duet:piscat,jones:piscat,ashery:pioncx,Bellotti1973_abs:piscat,Bellotti1973_cx:piscat} and ABS (red) and CX (black). This figure is taken from Ref.~\cite{Elder:2019}.}
  \label{fig:PiFSI-models}
\end{figure}

In total six types of pion scattering channels are defined: 
low momentum pion quasi-elastic scattering, 
low momentum pion charge exchange interaction,
pion absorption, 
high momentum pion quasi-elastic scattering, 
high momentum pion charge exchange interaction,
high momentum pion inelastic scattering (pion production).
Energy-independent normalization factors were defined for
each of these six channels and fit to pion--nucleus scattering data from various 
experiments by Pinzon Guerra~\emph{et al.}~\cite{Elder:2019}, covering pion energies up to 2GeV.
Fig. \ref{fig:PiFSI-models} shows the comparison between the
pion scattering data and the simulated results from \neut
and the other simulation.
The kinematics of particles after the interaction are determined
using the results of the phase shift analysis~\cite{Rowe:1978} 
with medium corrections as suggested by Seki~\emph{et al.}~\cite{Seki:1983}.

The interactions of kaons, etas and omegas are treated similarly
as high momentum pions. The mean free paths are provided as
functions of momentum for each meson and the local Fermi-gas
model is used to determine the kinematics when an
interaction occurs. 

For kaons, (quasi-)elastic
scattering and charge exchange interactions are simulated.
The mean free paths of these kaon interactions
are extracted from the results of the phase shift analysis of
kaon scattering data~\cite{Hyslop:1992cs,Martin:1977me}.
These results are also used to determine the kinematics of
 outgoing particles. 

For etas, the 
$\eta N \rightarrow
\eta N, \eta N \rightarrow \pi N'$, and
$\eta N \rightarrow \pi \pi N'$ processes are implemented. These interactions
are simulated assuming that an eta produces an excited nucleon state, which then decays to give final state. The  $N(1540), N(1650)$ intermediate
baryon resonances are considered. The 
production cross-section of baryon resonances are given by
the Breit-Wigner formula:
\begin{equation}
  \sigma(k) = \frac{\pi}{k^2}\cdot
  \frac{\Gamma_{\eta N}\Gamma_{X}}{(W-M_N^*)^2+\Gamma^2_{tot}/4},
\end{equation}
where $W$ and $M_N^*$ are the invariant mass of the intermediate
baryon resonance and the mean mass of the resonance respectively,
$\Gamma_{tot}$ is the total width of $N^*$,
$\Gamma_{\eta N}$ is the partial width of $N^* \rightarrow \eta N$,
$\Gamma_{X}$ is the partial width to the final state $X$, where
$X$ is the final state meson, $\eta$, $\pi$ or $\pi \pi$.
The produced particles are ejected isotropically
in the intermediate resonance rest frame.

For omegas, the 
$\omega N \rightarrow \pi N,
\omega N \rightarrow \rho N, 
\omega N \rightarrow \rho \pi N', 
\omega N \rightarrow \rho \pi \pi N',
\omega N \rightarrow \omega N,$ 
and
$\omega N \rightarrow \sigma(f_0) N'$ processes are considered. The total and 
differential cross-sections of each channel were calculated
following the prescription by Lykasov~\emph{et al.}~\cite{Lykasov:1998ma}.

\subsubsection{Nucleon interactions in nucleus}\label{sec:HadronInteraction:Nucleon}

\begin{figure}
  \centering
  \includegraphics[width=1\textwidth]{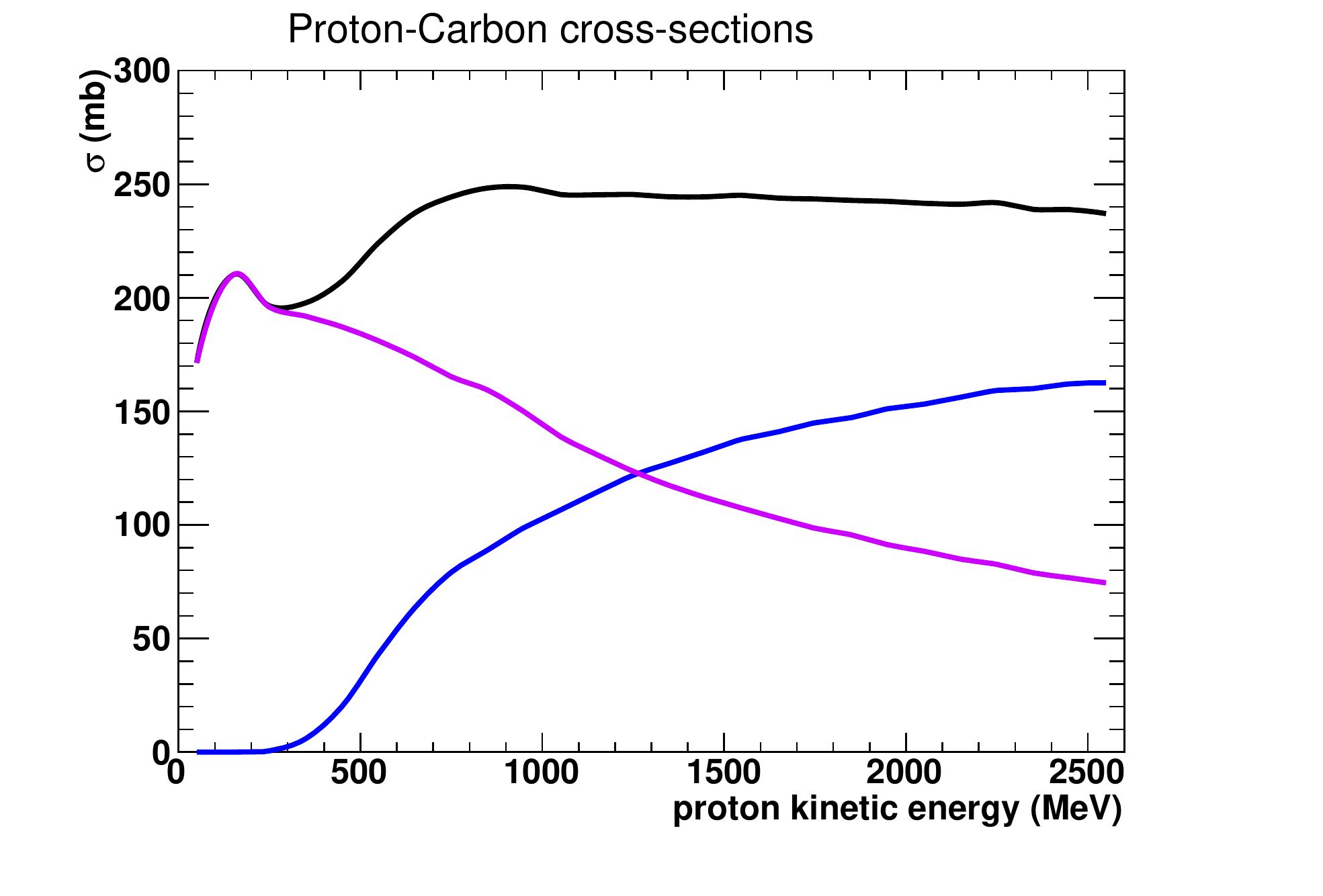}
  \caption{Proton-carbon scattering cross-sections as a function of the proton kinetic energy, as predicted by \neut. To obtain this plot, protons are injected from outside of the nucleus and \neut simulates the interactions in the carbon nucleus. The black line shows the total cross-section, the purple line shows the elastic scattering cross-section and the blue line shows the inelastic scattering cross-sections.}
  \label{fig:NucFSI-pC-xsec}
\end{figure}

The implementation of nucleon scattering is also based on 
the INC model. Three types of interaction are 
considered, elastic scattering and one or two pion production.
These are implemented following the work by 
Bertini~\emph{et al.}~\cite{Bertini:1972vz}, for MECC-7.
The same nuclear density function and local Fermi-gas model is used as the meson 
interactions in nucleus. When calculating the interaction kinematics, an effective nucleon mass ($M_N^{eff}$) is used instead of the free mass ($M_N^{free}$). Here, $M_N^{eff}$ is defined 
as 
\begin{equation}
  M_N^{eff} = \sqrt{(M_N^{free} - 8MeVc^2)^2-(p_F^{surf})^2},
\end{equation}
where $P_F^{surf}$ is the density-dependent local Fermi
surface momentum.
The momentum of the nucleon after an interaction
is required to be larger than the local $P_F^{surf}$, which is the
standard Pauli-blocking procedure in the local Fermi-gas model framework.
Produced pions are assumed to arise from the decay of 
$\Delta(1232)$ resonances produced by nucleon-
nucleon scattering. To simulate the $\Delta(1232)$ production 
a simple isobar model~\cite{Lindenbaum:1957ec} is used.
When a pion is produced in the nucleus after the nucleon
re-scattering, that pion is independently tracked from the 
point of generation using the pion transport simulation described in section~\ref{sec:HadronInteraction:meson}.
The predicted cross-sections for proton-carbon scattering is shown
in Fig. \ref{fig:NucFSI-pC-xsec}.

\subsection{Known Implementation Limitations}

\neut is primarily developed to enable world-leading \sk and \ttwok neutrino oscillation, proton decay, and neutrino cross-section measurements, for which a fully consistent description of Nature is desirable but not necessary, and as such has some known limitations beyond the imperfect predictions of the implemented physics models. 

The most prominent limitation is that the modelling of the initial nuclear state is tied to the individual interaction channel being simulated. For most channels, the initial state is always modelled as a Fermi-gas, for multi-nucleon scattering an LFG is used, and only for quasi-elastic scattering is the more sophisticated spectral function an option. The SF is only designed to model QE interactions, and only for specific nuclei where relevant electron-scattering data exist, for other nuclear targets a Fermi-gas is used. While, as noted above, the modelling of hadron re-scattering is based on the density and momentum predictions from an LFG model. Such an inconsistent model is sometimes affectionately referred to as a \emph{Franken-model}, after the fictional scientist and his Gothic horror implementation. For single meson production, nuclear effects beyond the initial-state nuclear momentum and Pauli-blocking are ignored. It might be expected that the inclusion of the initial-state nucleon removal energy would affect the predictions. As statistical uncertainties are reduced with the next generation of long-baseline oscillation experiments, addressing these inconsistencies will become a focus of future development. 

\neut does not attempt to model a number of interaction channels that are irrelevant to \sk and \ttwok analyses, these include neutrino--electron elastic scattering (used to provide complementary neutrino flux normalization constraints for higher energy beams~\cite{PhysRevD.100.092001}) and inverse beta decay (important for simulating reactor neutrino experiments).

\section{Additional Tools}\label{sec:Tools}

So far we have focused on the implementation and physics models for simulation single neutrino--nucleus interactions. While this constitutes the core of the simulation, to be an effective tool for data analysis, \neut provides a number of other tools and features.

\paragraph{\texttt{neutgeom}:} The \texttt{neutgeom} tool is used to interface to a neutrino beamline simulation, perform neutrino ray-tracing, interface with a detector geometry description and correctly distribute interaction positions through the detector. Currently, \texttt{neutgeom} is only able to consume neutrino flux \emph{vectors} from \texttt{JNUBEAM}, which simulation the J-PARC neutrino beam for the \ttwok experiment. There is some interest in distributing \texttt{neutgeom} as a standalone tool capable of interfacing with other neutrino beam simulations and interaction simulations.

\paragraph{\texttt{NReWeight}:} Cross section \emph{reweighting} is an important tool for systematic error estimation for neutrino-scattering analyses. For systematic parameters that can be effectively reweighted it enables interaction model variations to be applied at analysis time, rather than requiring re-simulation and re-analysis, reducing the computing time taken to estimate systematic uncertainties by many many orders of magnitude. At its core, the reweighting technique calculates a weight for each already-simulated interaction, 
\[
W = \frac{\sigma\left(p^\prime,\vec{x}\right)}{\sigma\left(p,\vec{x}\right)},
\]
where $\vec{x}$ encapsulates the kinematics of the simulated interaction and $p$ and $p^\prime$ fully describe the model choices and any free parameters in the already-simulated and the varied model, respectively. This procedure is only exact when the varied model, $p^\prime$, predicts a range of $\vec{x}$ that is the same or a strict subset of the range predicted by $p$. \neut provides the \texttt{NReWeight} package that implements exact cross section reweighting enabling the variation of nucleon form factors for CCQE and single meson production parameters after simulation. \texttt{NReWeight} also exposes reweighting for the meson and nucleon intranuclear cascade, which is exact for modest variations of the underlying meson--nucleon and nucleon--nucleon scattering probabilities. \texttt{NReWeight} is a critical tool for \sk and \ttwok analyses.

\paragraph{Interface to \texttt{GEANT3} and \texttt{GEANT4}:} The simulation of an intranuclear cascade and of meson--nuclear and nucleon--nuclear scattering is conceptually similar, the difference is whether the probe beings the simulation inside or outside of the simulated nucleus. As described in section~\ref{sec:HadronInteraction:meson}, the pion--nucleon interaction cross-sections were tuned to pion--nuclear scattering data. As a result, it is attractive for physics consistency to be able to simulate pion--nucleus scattering with the \neut INC model. To achieve this, an interface between \neut and \texttt{GEANT} was developed and integrated to the \ttwok near and SK detector simulation programs.
\section{Future direction}\label{sec:Direction}
Current and future neutrino experiments require precise neutrino interaction simulations to achieve their ambitious physics goals.
To meet these requirements, \neut will be continue to be developed and improved.

Current modelling improvements include the implementation of the state-of-the-art CCQE and multi-nucleon model by Amaro~\emph{et al.}~\cite{Amaro:2021sec},
and single pion production models by Kabirnezhad~\cite{Kabirnezhad:2020wtp}
and by Sato~\emph{et al.}~\cite{Sobczyk:2018ghy}.The implementation of a rudimentary QE-only electron-scattering simulation in \neut is underway. This will enable improved extended validations of the implemented physics that is most critical to \ttwok analyses.

The dependence on \cernlib is problematic, as the library is no longer maintained. Building and distributing the library for modern compilers and operating systems takes time away from more important development efforts. \neut currently relies on \cernlib for reading configuration files, random number generation and some common mathematical operations, and PYTHIA v5.72 for simulating SIS and DIS interactions. Resolving the dependence on PYTHIA v5 is not simple as changing to a newer version (v6 or v8) would require the implementation of the particles' kinematics determination (vector generation) functions in \neut, as the PYTHIA implementation that we rely on has been dropped or
does not cover the entire kinematic region. 
Removing the dependency on \cernlib is a high priority for near-future maintenance.

The current closed source nature of \neut is undesirable. Exposure to more users and use cases will result in code, interface, and physics improvements. However, the lack of human resources render it difficult to support \neut as a more general tool. Work has begun, in collaboration with other neutrino interaction simulation stake-holders, to define, test, and implement a new community-designed event format and event generation API~\cite{barrow2020summary}. 
These critical future developments will be implemented in \neut as they become defined and mature.

\section{Summary}
\neut is a general purpose neutrino interaction simulation used and improved by members of the \sk and the \ttwok collaborations, with critical additional contributions from interaction theory groups. Future development will target the physics requirements of Super-Kamiokande, T2K, and Hyper-Kamiokande, and software integration improvements to the \neut API and data formats.

\section{Acknowledgments}
The authors would like to thank the many contributors of \neut, and the users in
the SK, T2K, Ninja, and WAGASCI collaborations that push us to continue to develop it as a world-class analysis tool. The authors would also like to thank Kendall Mahn for help proofreading this document.
This work is partially supported by MEXT/JSPS KAKENHI Grant Number 18H05536. This material is based upon work supported by the U.S. Department of Energy, Office of Science, Office under Award Number(s) DE-SC0015903.

%

\end{document}